\documentclass[12pt]{article}

\usepackage[ round,authoryear]{natbib}

\usepackage{siunitx}
\usepackage{color}
\usepackage{amsbsy}
\usepackage{latexsym}
\usepackage{graphicx}
\usepackage{url}
\usepackage[colorlinks,linkcolor={blue},urlcolor={blue}, citecolor={blue}, filecolor={black}]{hyperref}
\usepackage{rotating}
\usepackage{multirow,caption,booktabs,float,mathrsfs}
\usepackage{authblk,enumitem,setspace}
\usepackage{amsthm,amsmath,amsfonts,amssymb}

\makeatletter
\@addtoreset{equation}{section}

\theoremstyle{plain}
\newtheorem{theorem}{Theorem}

\newtheorem{corollary}{Corollary}
\newtheorem{remark}{Remark}[section]

\newtheorem{proposition}{Proposition}
\newtheorem{definition}{Definition}

\def\cal{\mathcal}

\def\tilde{\widetilde}

\def\conP{\stackrel{\mathrm{P}} {\to}}                 
\def\conD{\stackrel{\cal D} {\to}}

\def\cdf{{\mathrm{c.d.f.}}}
\def\P{\,\mathbb{P}}
\def\I{\,\mathrm{I}}
\def\iid{\mathrm{i.i.d.}}
\def\E{{\mathbb{E}}}

\def\d{{\mathrm{d}}}
\def\wpone{\mathrm{a.s.}}  

\def\calT{\mathcal{T}}

\def\mathbbR{\mathbb{R}}

\def \var{\mathrm{var}}

\begin{document}

\title
{\bf
A family of Chatterjee's correlation coefficients and their properties
}
\author[1]{Muhong Gao}
\author[1,2]{Qizhai Li}

\affil[1]{\small Academy of Mathematics and System Science, Chinese Academy of Sciences}
\affil[2]{\small University of Chinese Academy of Sciences}

\date{\today}
\maketitle

\begin{abstract}
Quantifying the strength of functional dependence between random scalars $X$ and $Y$ is an important statistical problem. While many existing correlation coefficients excel in identifying linear or monotone functional dependence, they fall short in capturing general non-monotone functional relationships. In response, we propose a family of correlation coefficients $\xi^{(h,F)}_n$, characterized by a continuous bivariate function $h$ and a $\cdf$ function $F$. By offering a range of selections for $h$ and $F$,
$\xi^{(h,F)}_n$ encompasses a diverse class of novel correlation coefficients, while also incorporates the Chatterjee's correlation coefficient \citep{Chatterjee_2021} as a special case. We prove that  $\xi^{(h,F)}_n$ converges almost surely to a deterministic limit $\xi^{(h,F)}$ as sample size $n$ approaches infinity. In addition, under appropriate conditions imposed on
$h$ and $F$, the limit $\xi^{(h,F)}$ satisfies the three appealing properties: (P1). it belongs to the range of $[0,1]$; (P2). it equals $1$ if and only if $Y$ is a measurable function of $X$; and (P3). it equals $0$
if and only if $Y$ is independent of $X$. As amplified by our numerical experiments, our proposals provide practitioners with a variety of options to choose the most suitable correlation coefficient tailored to their specific practical needs. 
\end{abstract}


\section{Introduction}
Measuring the strength of functional dependence between two random scalars $X$ and $Y$ 
is a fundamental problem  in statistics with broad scientific implications.
In explicit statistical terms, the objective is to formulate a straightforward dependence estimate $\xi(X,Y)$, wherein a higher value of $\xi(X,Y)$ indicates a stronger dependence of $Y$ on $X$. Ideally, $\xi(X,Y)$ attains its maximum value if and only if $Y = f(X)$, where $f$ represents an arbitrary non-random measurable function.
Across the extensive literature spanning the past century that focuses on quantifying pairwise associations, the majority of proposals have been primarily crafted for identifying linear/monotonic dependence patterns or testing independence,
including the classical Pearson's correlation coefficient, Spearman's $\rho$, and Kendall's $\tau$, as well as the distance correlations \citep{Szekely_etal_2007_AOS, szekely2009brownian, Wang_etal_2015_JASA}, rank correlations \citep{drton2020high, Weihs_etal_2018_Biometrika, Bergsma_etal_2014_Bernoulli, Shi_Hallin_Drton_Han_2022_AOS}, copula-based correlations \citep{Schweizer_Wolff_1981,chang2016robust,lopez2013randomized, zhang2019bet}, binning-based correlations \citep{Heller_Heller_Gorfine_2012,Kinney_etal_2014_PNAS,Ma_Mao_2019_JASA}, among many others. While for measuring the amount of general functional dependence where $f$ is possibly to be non-monotone,  these approaches typically fall short in delivering meaningful and efficacious assessments.

\subsection{Chatterjee's correlation coefficient}
The recent genius work of  \cite{Chatterjee_2021} has attracted very much attention. It proposed a new elegant correlation coefficient that effectively measures the extent of functional dependence under arbitrary $f$ functions. Given the $\iid$ data copies $(X_1,Y_1),\dots, (X_n,Y_n) \sim (X,Y)$ with $n \geq 2$, 
the Chatterjee's correlation coefficient has a very simple expression:
\begin{eqnarray}
\xi_n(X,Y) = 1- \frac{\sum_{i=1}^{n-1}|R(Y_{[i+1]}) - R(Y_{[i]})|  }{\chi_n(Y)},  \label{eq:CCC}
\end{eqnarray}
where $(X_{(1)}, Y_{[1]}),\dots, (X_{(n)}, Y_{[n]})$ are reordered pairs of data such that $X_{(1)}\leq\dots \leq X_{(n)}$, and $R(Y_{[i]}) = \sum_{j=1}^n\I(Y_j \leq Y_{[i]})$ represents the rank of $Y_{[i]}$ among all $Y_j$'s. The denominator term $\chi_n(Y)$, acting as a normalization statistic, equals $(n^2-1)/3$ if there are no ties among $X_i$'s, and has a slightly altered expression in the presence of ties among $X_i$'s. 
Despite its simple non-parametric form, the most compelling feature of Chatterjee's correlation coefficient is its almost-surely convergence to a limit $\xi = \lim_{n\to \infty} \xi_n(X,Y) $,
which satisfies the following three desirable properties: 
\begin{enumerate}[label=\rm{(P\arabic*)}]
\item \label{CP_1} (Normalization).  $\xi \in [0,1]$, for all distributions of $(X,Y)$.
\item \label{CP_2} $($Perfect dependence$)$. $\xi =1 $ (reaches its maximum) if and only if there is a measurable function $f : \mathbbR \to \mathbbR$, such that $Y=f(X)$ almost surely. 
\item \label{CP_3} $($D-Consistency$)$. $\xi =0$ (reaches its minimum) if and only if $X$ and $Y$ are independent. 
\end{enumerate}
These three properties of $\xi$, as first introduced by a prior work \citep{Dette_etal_2013}, appear to be so strong  that none of the preceding correlation measures can simultaneously fulfill all of them.
In particular, the perfect dependence property \ref{CP_2} 
stands as a distinguished characteristic of Chatterjee's correlation coefficient,
granting $\xi_n(X,Y)$ extraordinary capability in reflecting functional dependence.
Several other correlation coefficients share the property of being maximized when $Y=f(X)$ is satisfied, including the maximal correlation coefficient \citep{Renyi_1959}, the maximum information coefficient \citep{Reshef_etal_2011_Science}, and the Hellinger correlation \citep{Geenens_etal_2022_JASA}. But the converse is not true: they may also achieve maximum values when $Y$ and $X$ exhibit noisy or non-functional relationships. Consequently, their identification of functional dependence can sometimes be inaccurate and misleading.

\subsection{Motivation}
We first present an intuitive rationale for the efficacy of Chatterjee's correlation coefficient \eqref{eq:CCC}. 
One ingenuous aspect of \eqref{eq:CCC}  resides in the formulation of the numerator term $\sum_{i=1}^{n-1}|R(Y_{[i+1]}) - R(Y_{[i]})|$. 
Here, $|R(Y_{[i+1]}) - R(Y_{[i]})|$ is noted for quantifying the degree of variation between two consecutive $Y_{[i]}$'s. Considering that the $(X_{(i)}, Y_{[i]})$ pairs are rearranged based on the order of $X_{(i)}$ values, the variation between two consecutive $Y_{[i]}$ and $Y_{[i+1]}$ tends to be small when the dependence of $Y$ on $X$ is significant. Conversely, a greater variation is likely to occur when $Y$ exhibits greater independence from $X$. Therefore, achieving the maximum or minimum of \eqref{eq:CCC} fundamentally relies on the dependence level between $Y$ and $X$. 

This motivates a natural conjecture regarding potential extensions of Chatterjee's correlation coefficient: replacing the term $|R(Y_{[i+1]}) - R(Y_{[i]})|$ in \eqref{eq:CCC} by other types of metrics on the degree of variation between $Y_{[i]}$ and $Y_{[i+1]}$ may still be effective for measuring functional dependence. 
This paper is dedicated to conducting a detailed investigation into this interesting question.

\subsection{Our contributions} \label{Sec_1.3}
We propose a new correlation coefficient
\begin{eqnarray}
\xi^{(h,F)}_n(X,Y) = 1- \frac{\sum_{i=1}^{n-1} h\big(F(Y_{[i]}), F(Y_{[i+1]})\big) }{\chi^{(h,F)}_n(Y)}, \label{eq:our_CCC}
\end{eqnarray}
with $F(\cdot)$ being a monotone univariate function, and $h(\cdot,\cdot)$ being a non-negative continuous bivariate function, satisfying $h(x,x)=0$ for $x \in \mathbbR$. To enhance identifiability, we further restrict $F(\cdot)$ to be a cumulative distribution function ($\cdf$). The denominator $\chi^{(h,F)}_n(Y)$ serves as a normalization statistic,
whose explicit expression will be presented later in Section \ref{Section_2+}. 
Comparing \eqref{eq:our_CCC} with \eqref{eq:CCC}, the term $h(F(Y_{[i]}), F(Y_{[i+1]}))$ in \eqref{eq:our_CCC} is an extension of the corresponding term $|R(Y_{[i+1]}) - R(Y_{[i]})|$ in \eqref{eq:CCC}, capable for quantifying the degree of variation between $Y_{[i]}$ and $Y_{[i+1]}$. 
With a variety of choices for $h$ and $F$ functions, $\xi^{(h,F)}_n(X,Y)$ encompasses a wide family of new correlation coefficients that have not been previously explored in existing literature. 
It is noticeable that the Chatterjee's correlation coefficient \eqref{eq:CCC} corresponds to a special case  of $\xi^{(h,F)}_n(X,Y)$ in \eqref{eq:our_CCC} when taking $h(x,y)= n\cdot|x-y|$, and $F(x) = F_{Y,n}(x) = 1/n\cdot\sum_{i=1}^{n} \I(Y_i\leq x)$ as the empirical $\cdf$ of $Y_i$'s. 

This paper seeks to provide answers to the fundamental questions concerning the asymptotic behaviors and limit properties of $\xi^{(h,F)}_n(X,Y)$, which can be contingent upon the selection of $h$ and $F$. We obtain a series of valuable results as outlined below. 
\begin{enumerate}
\item $\xi^{(h,F)}_n(X,Y)$ converges almost surely to a limit $\xi^{(h,F)}$. 
\item The limit $\xi^{(h,F)}$ satisfies the perfect dependence property \ref{CP_2}, under mild regularity conditions on $h$ and $F$. 
\item The limit $\xi^{(h,F)}=0$ as long as $X$ and $Y$ are independent. Moreover, a central limit theorem is established for $\xi^{(h,F)}_n(X,Y)$ in the case where $X$ and $Y$ are independent. 
\item  
We establish the necessary and sufficient condition for $\xi^{(h,F)}$ to satisfy all properties \ref{CP_1}--\ref{CP_3}. Specifically, we delineate the set $\mathcal{H}$ comprising continuous bivariate functions, wherein $\xi^{(h,F)}$ possesses properties \ref{CP_1}--\ref{CP_3} if and only if $h \in \mathcal{H}$. Our findings reveal that the set $\mathcal{H}$ is extensive, encompassing a diverse array of common bivariate functions.
\item While the coefficient $\xi^{(h,F)}_n(X,Y)$ is initially formulated using deterministic $h$ and $F$ functions, 
we also allow taking $F=F_{Y,n}$ as the empirical $\cdf$ of $Y_i$'s. 
In this scenario, $\xi^{(h,F_{Y,n})}_n(X,Y)$ becomes a rank-based correlation coefficient, for which the majority of the above four results are still valid. 
\end{enumerate}

The above results admit the existence of a wide family of new correlation coefficients that have theoretically-backed capability in measuring functional dependence between bivariate random variables. 
When dealing with real-world problems, 
our $\xi^{(h,F)}_n(X,Y)$ can better cater to practical needs by adapting $h$ and $F$ functions in a manner tailored to specific purposes. This flexibility contrasts with the constrained structure of the Chatterjee's correlation coefficient.
Extensive numerical experiments in Sections \ref{Sec_5} and \ref{Sec_6} lend empirical support to the practical efficacy of our proposals. 
\paragraph*{Outline of the paper.} Section \ref{Section_2+} formally presents our correlation coefficient $\xi^{(h,F)}_n(X,Y)$  alongside its fundamental asymptotic properties. Section \ref{Sec_3} demonstrate the necessary and sufficient condition for achieving all desired properties \ref{CP_1}--\ref{CP_3}. Section \ref{Sec_4} investigate the properties of the rank-based variant of $\xi^{(h,F)}_n(X,Y)$ when taking $F = F_{Y,n}$. Section \ref{Sec_5} conducts extensive simulations for accessing the proposed correlation coefficient. Section \ref{Sec_6} showcases an application of our correlation coefficient on a real-world dataset. Section \ref{Sec_8} briefly concludes. Proofs of theorems are collected in the Supplementary Material. 

\paragraph*{Notations.} For integer $n \geq 2$, let $(X_1,Y_1),\dots, (X_n,Y_n)$ be $\iid$ samples of $(X,Y)$ in the probability space $\{\Omega, \mathscr{F}, \P\}$. Denote by $(X_{(1)}, Y_{[1]}),\dots, (X_{(n)}, Y_{[n]})$ the reordered pairs of samples such that $X_{(1)}\leq\dots \leq X_{(n)}$. If there are ties among $X_i$'s, breaking the ties uniformly at random in this ordering. 
Let $\I(\cdot)$ denote the indicator operator. 
Let $F_{X}(t) :=  \P(X \leq t)$ and $F_{Y}(t) :=  \P(Y \leq t)$ be the $\cdf$'s of $X$ and $Y$ respectively, and let $F_{Y,n}(t) = 1/n\cdot\sum_{i=1}^{n} \I(Y_i\leq t)$ be the empirical $\cdf$ of $Y_i$'s. Denote by $\Phi(\cdot)$ the c.d.f of standard normal distribution. 
Denote by $\mu_{X}$ and $\mu_{Y}$ the distribution laws of $X$ and $Y$. Let $\mu_{Y|X=x}$ (or $\mu_{Y|x}$ for short) denote the conditional law of $Y$ given $X=x$. For the existence of $\mu_{Y|x}$, refer to Theorem 2.1.22 and Exercise 4.1.18 in \cite{Durrett_2019} (also see Lemma 1 in the Supplementary Material). Let $\conP$ and $\conD$ denote convergence in probability and in distribution respectively. For simplicity, the term ``almost surely'' is abbreviated as ``$\wpone$'' in case needed.

\section{The new family of correlation coefficients} \label{Section_2+}
In this section, we formally present the formulation of our proposed correlation coefficient along with its fundamental properties.  Throughout the paper, we assume that neither $X$ nor $Y$ is a constant. 

\begin{definition}[New correlation coefficient] \label{Def_1}
For a non-negative continuous bivariate function $h: [0,1]^2 \to [0,\infty)$ satisfying $h(x,x) =0$ for all $x \in [0,1]$, and a $\cdf$ function $F:\mathbbR \to [0,1]$, 
define the variation statistic
\begin{eqnarray*} 
\zeta^{(h,F)}_n(X,Y) := \frac{1}{n}\sum_{i=1}^{n-1} h\big( F(Y_{[i]}), F(Y_{[i+1]})\big), 
\end{eqnarray*} and the normalization statistic 
\begin{eqnarray*}
\chi^{(h,F)}_n(Y) := \frac{1}{n^2}\sum_{i=1}^n \sum_{j=1}^n h(F(Y_i), F(Y_j)). 
\end{eqnarray*} The new correlation coefficient is defined as:
\begin{eqnarray*}
\xi^{(h,F)}_n(X,Y) &:=& 1 - \zeta^{(h,F)}_n(X,Y)/\chi^{(h,F)}_n(Y) \cr
&=& 1 - \frac{n \cdot\sum_{i=1}^{n-1} h\big( F(Y_{[i]}), F(Y_{[i+1]})\big)}{\sum_{i=1}^n \sum_{j=1}^n h(F(Y_i), F(Y_j))}.
\end{eqnarray*}
In case that $\chi^{(h,F)}_n(Y) =0 $, we set $\xi^{(h,F)}_n(X,Y) = 1$. 
\end{definition}

As previously mentioned in Section \ref{Sec_1.3}, the numerator term	$\zeta^{(h,F)}_n(X,Y)$ is designed to measure the degree of variation between consecutive $Y_{[i]}$'s. The denominator term $\chi^{(h,F)}_n(Y)$  serves as a normalization statistic, ensuring that $	\xi^{(h,F)}_n(X,Y) $ tends towards zero once if $X$ and $Y$ are independent.
Theorem \ref{Thm_1} shows that $\xi^{(h,F)}_n(X,Y)$ converges almost surely to a deterministic constant $\xi^{(h,F)}$ as $n$ approaches infinity. 

\begin{theorem}[Convergence of correlation coefficient] \label{Thm_1}
As $n \to \infty$, we have the following three convergence results:

\noindent $\mathrm{(i)}$ the variation statistic $\zeta^{(h,F)}_n(X,Y)$ converges a.s. to the limit
\begin{eqnarray*}
\zeta^{(h,F)} &:=& \int\int\int h\big(F(y),F(z)\big)\ \d \mu_{Y|x}(y) \ \d \mu_{Y|x}(z) \ \d \mu_{X}(x);  
\end{eqnarray*}
$\mathrm{(ii)}$ the normalization statistic $\chi^{(h,F)}_n(Y) $ converges a.s. to the limit
\begin{eqnarray*}
\chi^{(h,F)} := \int\int h(F(y),F(z))\ \d \mu_{Y}(y) \ \d \mu_{Y}(z); 
\end{eqnarray*}
and $\mathrm{(iii)}$ the correlation coefficient $\xi^{(h,F)}_n(X,Y)$ converges a.s. to the limit 
\begin{eqnarray}
\xi^{(h,F)} &:=& 1- \zeta^{(h,F)} /\chi^{(h,F)}.     \label{eq: xi}
\end{eqnarray}
In case that $\chi^{(h,F)}=0$, we set $\xi^{(h,F)}=1$.
\end{theorem}

Throughout the paper, we call the population-level deterministic quantity $\xi^{(h,F)}$ the ``correlation measure'', to distinguish it from its consistent empirical estimator $\xi^{(h,F)}_n(X,Y)$ which is called the ``correlation coefficient''. 
Several remarks are provided in order. 

\begin{remark}
The new correlation coefficient $\xi^{(h,F)}_n(X,Y)$ 
requires no need for estimating probability densities, characteristic functions or mutual information, and it could be simply computed with a time complexity of $O(n^2)$.  
Moreover, $h$ and $F$ in $\xi^{(h,F)}_n(X,Y)$  are both prespecified deterministic functions, without the need of being tuned or estimated from the observed data. 
\end{remark}

\begin{remark}
In Theorem \ref{Thm_1}, there are no constraints on the distribution of $(X,Y)$ except that neither $X$ nor $Y$ is a constant. Thus Theorem \ref{Thm_1} is applicable to both discrete and continuous $(X,Y)$ distributions. 
\end{remark}

\begin{remark}
Theorem \ref{Thm_1} is almost assumption-free on $h$ and $F$, 
except that $h$ is required to be continuous (as specified in Definition \ref{Def_1}).
This continuity condition for $h$ is vital for establishing Theorem \ref{Thm_1}. 
Lemma A.6 in Appendix A of the Supplementary Material shows a trivial example of a discontinuous $h(y,z)= \I(y\neq z)$ for which  $\xi^{(h,F)}_n(X,Y)$ does not converge to $\xi^{(h,F)}$. 
\end{remark}

\begin{remark}
When taking $h(y,z) = |y-z|$ and $F = F_Y$ as the $\cdf$ of $Y$ in \eqref{eq: xi}, it could be verified (see Lemma A.7 in Appendix A of the Supplementary Material) that the resulting correlation measure $\xi^{(h,F_Y)}$  has an alternative expression
\begin{eqnarray*}
\xi^{(h,F_Y)} = 1- \frac{\zeta^{(h,F_Y)} }{\chi^{(h,F_Y)}} = \frac{\int\var\big(\E(\I(Y\geq t)\mid X)\big) \ \d \mu_{Y}(t)}{\int\var\big(\I(Y\geq t)\big) \ \d \mu_{Y}(t)},
\end{eqnarray*}
which coincides with the correlation measure $\xi$ in \cite{Chatterjee_2021} [Theorem 1.1, Eq. (2)].

\end{remark}

\begin{remark}
From our definition, it is noted that	$\xi^{(h,F)}_n(X,Y)$ or $\xi^{(h,F)}$ fully depends on the functions $h$ and $F$, yet this relationship is not ``identifiable''. In other words, 
for distinct pairs of functions $(h_1, F_1)$ and $(h_2,F_2)$, $\xi^{(h_1,F_1)}_n(X,Y)$ and $ \xi^{(h_2,F_2)}_n(X,Y)$ could be identical correlation coefficients, provided that the composite functions $h_1(F_1(\cdot), F_1(\cdot))$ and $h_2(F_2(\cdot), F_2(\cdot))$ are the same. 
However, 
treating $h$ and $F$ separately is intentional, because 
$h$ and $F$ have different influences on the properties of the correlation coefficient. Such design of separate $h$ and $F$ not only facilitates further theoretical derivations, but also provides more concrete guidelines for constructing the correlation coefficients in real practice. 
\end{remark}

\begin{remark} 
Note that the function $h$ in $\xi^{(h,F)}_n(X,Y)$ is not required to be symmetric, which provides more flexibility. 
While for non-symmetric $h$,
the correlation measure $\xi^{(h,F)}$ in \eqref{eq: xi} remains unchanged if replacing this $h$ with the symmetric $h^\dagger(y,z) := (h(y,z)+h(z,y))/2$. Thus, when it turns to study the properties of $\xi^{(h,F)}$, assuming $h$ to be symmetric will not lose generality. For convenience, we assume that $h$ is symmetric in the remainder of paper (refer to condition \ref{Ch1}).
\end{remark}

\subsection{Perfect dependence property}
For $\xi^{(h,F)}$ to serve as a reliable measure for the strength of functional dependence, 
one of the most desirable properties is the perfect dependence property \ref{CP_2}. Theorem \ref{Thm_2} verifies that $\xi^{(h,F)}$ indeed possesses property \ref{CP_2}, provided that $h$ and $F$ satisfy the regularity conditions below.
\begin{enumerate}[label=\rm{(C\arabic*)}]
\item \label{Ch1} (Regularity condition on $h$). $h: [0,1]^2 \to [0,\infty)$ is a non-negative continuous symmetric function, satisfying $h(x,x) = 0$, and $h(x,y)>0$ for all $x,y \in [0,1]$.
\item \label{CF1} (Regularity condition on $F$). $F: \mathbbR \to (0,1)$ is a continuous and strictly-increasing $\cdf$ function. 
\end{enumerate}

\begin{theorem}[Properties of correlation measure] \label{Thm_2}
Assume that $h$ and $F$ satisfy the regularity conditions \ref{Ch1} and \ref{CF1}. 
Then the correlation measure $\xi^{(h,F)}$ in \eqref{eq: xi} possesses the perfect dependence property \ref{CP_2}. Moreover,  $\xi^{(h,F)}$ also satisfies 
the I-consistency property \ref{CP_3+}: 
\begin{enumerate}[label=\rm{(P\arabic*)}$'$]
\addtocounter{enumi}{2}  
\item \label{CP_3+} $($I-Consistency$)$. If $X$ and $Y$ are independent, then $\xi^{(h,F)} =0$. 
\end{enumerate}
\end{theorem}

For the above results, we give some remarks. 

\begin{remark} \label{remark_6} The sufficiency part of property \ref{CP_2} in Theorem \ref{Thm_2} can be illustrated as follows. When $Y=f(X)$, the conditional probability measure $\mu_{Y|x}$ reduces to the Dirac measure $\delta_{f(x)}$ at point $f(x)$. Thus we have
\begin{eqnarray*}
\zeta^{(h,F)} &=& \int\int\int h\big(F(y),F(z)\big)\ \delta_{f(x)}(y) \ \d \delta_{f(x)}(z) \ \d \mu_{X}(x) \cr
&=& \int h\big(F(f(x)),F(f(x))\big) \ \d \mu_{X}(x)  \,= \, 0,
\end{eqnarray*} 
from which $\xi^{(h,F)}=1$ follows. 
\end{remark}

\begin{remark}
The main purpose for imposing the regularity conditions \ref{Ch1}and \ref{CF1}, is to ensure that $h(F(y), F(z))$ is non-zero for any distinct $y,z \in \mathbbR$. This further admits the necessity part of property \ref{CP_2} in Theorem \ref{Thm_2}. If there exists distinct $y,z$ such that $h(F(y), F(z))=0$, then any random variable $Y$ with the support $\{y,z\}$ would lead to $\zeta^{(h,F)} = \chi^{(h,F)} =0$ and $\xi^{(h,F)} =1$, which violates  property \ref{CP_2}. Note that condition \ref{CF1} is satisfied if $F$ is a $\cdf$ of some continuous random variable with support of $\mathbbR$. 
\end{remark}

\begin{remark} \label{remark_8}The property \ref{CP_3+} in Theorem \ref{Thm_2}, which is called the ``I-consistency'' property \citep{Weihs_etal_2018_Biometrika}, corresponds to the sufficiency part of the ``D-consistency'' property \ref{CP_3}. Its proof can be seen as follows. 
If $X$ and $Y$ are independent, then the conditional probability measure $\mu_{Y|x}$ becomes $\mu_{Y}$ for all $x \in \mathbbR$. This gives that  
\begin{eqnarray*}
\zeta^{(h,F)} &=&\int\int\int h\big(F(y),F(z)\big)\ \mu_{Y}(y) \ \d \mu_{Y}(z) \ \d \mu_{X}(x) \cr
&=& \int\int h\big(F(y),F(z)\big)\  \d  \mu_{Y}(y) \ \d \mu_{Y}(z)  \,= \, \chi^{(h,F)},
\end{eqnarray*}
which yields $\xi^{(h,F)} = 1- \zeta^{(h,F)} /\chi^{(h,F)} =0$. 
\end{remark}

\subsection{Central limit theorem}
Measuring the amount of functional dependence is the major motivation for devising the correlation coefficient $\xi^{(h,F)}_n(X, Y)$, while testing independence is not our initial focus. In particular, the correlation measure $\xi^{(h,F)}$ under the mild regularity conditions \ref{Ch1} and \ref{CF1} satisfies the I-consistency property \ref{CP_3+}, whereas the stronger D-consistency property \ref{CP_3} will be derived in the next Section \ref{Sec_3}, after imposing further restrictions on $h$. But still, we derive the central limit theorem for $\xi^{(h,F)}_n(X, Y)$ under independent $X$ and $Y$,
suggesting its potential utility for conducting independence tests if desired.

\begin{theorem}[CLT of $\xi^{(h,F)}_n(X,Y)$ under independence]\label{Thm_3}
Assume that $h$ and $F$ satisfy the regularity conditions \ref{Ch1} and \ref{CF1}.
Assume that $X$ and $Y$ are independent.  Then as $n \to \infty$, we have
\begin{eqnarray*}
\sqrt{n} \cdot \xi^{(h,F)}_n(X,Y) \conD N(0, \sigma^2), 
\end{eqnarray*}
with 
\begin{eqnarray*}
\sigma^2 =	\frac{\E h(F(Y_1),F(Y_2))^2 - 2\cdot\E \big( h(F(Y_1),F(Y_2))\cdot h(F(Y_1),F(Y_3)) \big) +\big(\E h(F(Y_1),F(Y_2))\big)^2 }{\big(\E h(F(Y_1),F(Y_2))\big)^2}
\end{eqnarray*}
being strictly-positive, 
where $Y_1,Y_2,Y_3$ are $\iid$ copies of $Y$. 
\end{theorem}

The asymptotic variance $\sigma^2$ in Theorem \ref{Thm_3} could be consistently estimated via constructing U-statistics on $Y_i$'s; see details in Lemma A.13 in Appendix A of the Supplementary Material. 
In the special case of $h(y,z) = |y-z|$, $F = F_Y$, and $Y$ being continuous, we have $\sigma^2 = 2/5$, which coincides with the variance $\tau^2$ in \cite{Chatterjee_2021} [Theorem 2.2]. Also see Theorem \ref{Thm_6} and Remark \ref{remark_4.1} in Section \ref{Sec_4} for more related discussions.

\section{Pursuing all properties \ref{CP_1}--\ref{CP_3}} \label{Sec_3}
Owning the perfect dependence property \ref{CP_2}, the family of correlation coefficients $\{\xi^{(h,F)}_n(X,Y):$
$h$ and $F$ satisfies conditions \ref{Ch1} and \ref{CF1}$\}$ are readily capable for measuring strength of dependence of $Y$ on $X$. To explore further understandings for the correlation coefficient, a natural question arises: for which cases of $h$ and $F$ does the corresponding correlation measure $\xi^{(h,F)}$ satisfy all the three desirable properties \ref{CP_1}--\ref{CP_3}? This section unveils the answer to this question.

\subsection{Pursuing property \ref{CP_1}}
We first inspect the ``normalization property''  \ref{CP_1}. From the definition of correlation measure $\xi^{(h,F)} =1- \zeta^{(h,F)} /\chi^{(h,F)}$ in \eqref{eq: xi}, achieving property \ref{CP_1} means that the difference 
$\chi^{(h,F)} -\zeta^{(h,F)}$ is non-negative under any distributions of $(X,Y)$.
Proposition \ref{Prop_1} shows that this is achieved if and only if an additional condition \ref{Ch2} is imposed on function $h$.

\begin{enumerate}[label=\rm{(C\arabic*)}]
\addtocounter{enumi}{2}  
\item (Equivalence condition on $h$ for achieving \ref{CP_1}). \label{Ch2} 
There exists a continuous positive-definite symmetric function $\varphi: [0,1]^2 \to \mathbbR$, such that
\begin{eqnarray}
h(y,z) = 1/2\cdot \{\varphi(y,y)+\varphi(z,z)\} -  \varphi(y,z), \qquad  \text{for } y,z \in [0,1]. \label{eq: Ch2}
\end{eqnarray}
\end{enumerate}
Here, we call a symmetric bivariate function $\varphi$ ``positive-definite'', if for any real numbers $y_1,\dots, y_m\in [0,1]$ and $\omega_1,\dots, \omega_m \in \mathbbR$ with $m \geq 1$, $\sum_{i=1}^m \sum_{j=1}^m \omega_i \omega_j\varphi(y_i,y_j) \geq 0$ holds. 

\begin{proposition}[Necessary and sufficient condition for property \ref{CP_1}] \label{Prop_1}
Assume that $h$ and $F$ satisfy the regularity conditions \ref{Ch1} and \ref{CF1}.
Then the correlation measure $\xi^{(h,F)}$ satisfies property \ref{CP_1} if and only if $h$ satisfies condition \ref{Ch2}.
\end{proposition}

\begin{remark} 
We provide some sketch of proof for Proposition \ref{Prop_1}.
A closer inspection on $\chi^{(h,F)}$ and $\zeta^{(h,F)}$ in
Theorem \ref{Thm_1} yields a useful identity:
$		\chi^{(h,F)} -\zeta^{(h,F)} = $
\begin{eqnarray*}
\int \left[\int\int -h\big(F(y),F(z)\big)\cdot \big\{\d \mu_{Y|x}(y) -\d \mu_Y(y)\big\} \cdot \big\{\d \mu_{Y|x}(z) -\d \mu_Y(z)\big\} \right]\d \mu_{X}(x).  \label{eq: zeta+}
\end{eqnarray*}
Note that the double integral in the square bracket is symmetric with respect to $y$ and $z$, with $-h\big(F(y),F(z)\big)$ being its integral kernel. Thus, any negative-definite function $h$ would guarantee the positiveness of this double integral. This partly reflects why $h$ in \eqref{eq: Ch2} should include a term ``$-\varphi(y,z)$'', which is negative-definite. After incorporating $-\varphi(y,z)$ into $h$, the purpose for adding the other two terms $\varphi(y,y)$ and $\varphi(z,z)$ is to ensure that $h$ is nonnegative and $h(y,y)=0$ is satisfied for all $y \in [0,1]$. For more details, refer to Lemmas B.2 and B.3 in Appendix B of the Supplementary Material.  
\end{remark}

\subsection{Pursuing properties \ref{CP_1}--\ref{CP_3}} \label{Sec_3.2}
Combining the results in Theorem  \ref{Thm_2} and Proposition \ref{Prop_1},
we have derived the necessary and sufficient condition such that $\xi^{(h,F)}$ satisfies properties \ref{CP_1}, \ref{CP_2}, and \ref{CP_3+}. 
As long as the normalization property  \ref{CP_1} and the ``I-consistency'' property \ref{CP_3+} are both achieved, $\xi^{(h,F)}$ is guaranteed to attain its minimum value of $0$ under independent $X$ and $Y$. To fulfill all the three desired properties \ref{CP_1}--\ref{CP_3}, the last step is to upgrade property \ref{CP_3+} to property \ref{CP_3}. That is, we just need to ensure that there exists no dependent $X$ and $Y$ such that  $\xi^{(h,F)} =0$.  This is done by adding an further condition \ref{Cphi1} on the positive-definite function $\varphi$ in \eqref{eq: Ch2}.

\begin{enumerate}[label=\rm{(C\arabic*)}]
\addtocounter{enumi}{3}  
\item \label{Cphi1} 
(Condition on $\varphi$). $\varphi: [0,1]^2 \to \mathbbR$ is a continuous positive-definite symmetric function, such that for any two distinct probability measures $\mu_1$ and $\mu_2$ satisfying $\mu_1((0,1)) = \mu_2((0,1))=1$, the set
$\calT_{\mu_1, \mu_2} := \{t \in [0,1]: \E_{Z\sim \mu_1} \varphi(t, Z) \neq \E_{Z\sim \mu_2} \varphi(t, Z) \}$ 	is non-empty. 
\end{enumerate}
Accordingly, condition \ref{Ch2} of $h$ is upgraded to the following condition \ref{Ch3}.
\begin{enumerate}[label=\rm{(C\arabic*)}]
\addtocounter{enumi}{4}  
\item (Equivalence condition on $h$ for achieving \ref{CP_1}--\ref{CP_3}). \label{Ch3} 
There exists a bivariate function $\varphi: [0,1]^2 \to \mathbbR$ satisfying condition \ref{Cphi1}, such that $h(y,z) = 1/2\cdot \{\varphi(y,y)+\varphi(z,z)\} -  \varphi(y,z)$ holds for all $y,z \in [0,1]$. 
\end{enumerate}
Theorem \ref{Thm_4} verifies that condition \ref{Ch3} is the necessary and sufficient condition on $h$ such that $\zeta^{(h,F)}$ achieves all properties  \ref{CP_1}--\ref{CP_3}.
\begin{theorem}[Necessary and sufficient condition for all properties \ref{CP_1}--\ref{CP_3}] \label{Thm_4}
Assume that $h$ and $F$ satisfy the regularity conditions \ref{Ch1} and \ref{CF1}.
Then the correlation measure $\xi^{(h,F)}$ satisfies properties \ref{CP_1}, \ref{CP_2}, and \ref{CP_3}, if and only if $h$ satisfies condition \ref{Ch3}.
\end{theorem}

To satisfy condition \ref{Cphi1}, $\varphi$ needs to be a positive-definite function such that the mapping from a probability law $\mu$ to the univariate function $\E_{Z\sim \mu} \varphi(\cdot, Z)$ is one-to-one. 
This is equivalent to saying that 
$\varphi$ is a characteristic kernel function \citep{Sriperumbudur_etal_2011}, as defined in the context of the reproducing kernel Hilbert space. 
Examples of $\varphi$ satisfying condition \ref{Cphi1} includes: 
\begin{enumerate}[label=(\alph*)]
\item 
$\varphi(y,z) =  |y|^\gamma + |z|^\gamma - |y-z|^\gamma$, for $\gamma \in (0,2)$, which gives $h(y,z) = |y-z|^\gamma$; 
\item $\varphi(y,z) = \exp(-\beta \cdot |y-z|)$, for $\beta >0$, which yields $h(y,z) = 1 -  \exp(-\beta \cdot |y-z|)$.
\end{enumerate}
Refer to \cite{fukumizu2008characteristic} for a variety of other $\varphi$ functions that satisfy condition \ref{Cphi1}.

To provide a concise overview, Corollary \ref{Coro_1} summarizes the
results in the previous Theorems \ref{Thm_2}, \ref{Thm_4}, and Proposition \ref{Prop_1}. 
\begin{corollary} [Summary of  results] \label{Coro_1}
Assume that $h$ and $F$ satisfy the regularity conditions \ref{Ch1} and \ref{CF1}. We have the following results. 
\begin{enumerate}[label=\rm{(\roman*)}]
\item $\xi^{(h,F)}$ possesses properties \ref{CP_2} and \ref{CP_3+}.
\item $\xi^{(h,F)}$ possesses properties \ref{CP_1}, \ref{CP_2} and \ref{CP_3+}, if and only if $h$ satisfies condition \ref{Ch2}.
\item $\xi^{(h,F)}$ possesses properties \ref{CP_1}, \ref{CP_2} and \ref{CP_3}, if and only if $h$ satisfies condition \ref{Ch3}.
\end{enumerate}
\end{corollary}

\begin{remark}
To illustrate each result in Corollary \ref{Coro_1}, we showcase the properties of $\xi^{(h,F)}$ with $h(y,z) = |y-z|^\gamma$, for  $\gamma\in (0,\infty)$.  
Case (1): $\gamma\in (0,2)$. Then $h$ satisfies condition \ref{Ch3} with $\varphi(y,z) = |y|^\gamma + |z|^\gamma - |y-z|^\gamma$ satisfying condition \ref{Cphi1}. Thus, $\xi^{(h,F)}$  possesses all properties \ref{CP_1}, \ref{CP_2} and \ref{CP_3}. Case (2): $\gamma= 2$. Then $h$ satisfies condition \ref{Ch2}, with $\varphi(y,z) = y \cdot z$. Since this $\varphi$ does not satisfy condition \ref{Cphi1}, $\xi^{(h,F)}$  possesses properties \ref{CP_1}, \ref{CP_2} and \ref{CP_3+}, but violates \ref{CP_3}. Case (3): $\gamma\in (2,\infty)$. Then $h$ does not satisfy condition \ref{CP_1}, because $\varphi(y,z) = |y|^\gamma + |z|^\gamma - |y-z|^\gamma$ with $\gamma\in (2,\infty)$ is not positive-definite. Thus, $\xi^{(h,F)}$ possesses properties \ref{CP_2} and \ref{CP_3+}, but violates \ref{CP_1} and \ref{CP_3}. 
\end{remark}

\section{Rank-based correlation coefficients} \label{Sec_4}
As previously noted, the rank-based Chatterjee's correlation coefficient \eqref{eq:CCC} corresponds to our coefficient $\xi^{(h,F)}_n(X,Y) $ when taking $h(y,z) = |y-z|$ and $F = F_{Y,n}$ as the empirical $\cdf$ of $Y$. This motivates us to
investigate the rank-based correlation coefficient $\tilde \xi^{(h)}_n(X,Y)$ in Definition \ref{Def_2+}, which is a variant of the previous $\xi^{(h,F)}_n(X,Y) $ when letting $F= F_{Y,n}$. 
\begin{definition}[Rank-based correlation coefficient] \label{Def_2+}
For $y \in \mathbbR$, let $R(y) = n\cdot F_{Y,n} (y) = \sum_{j=1}^n \I(Y_j \leq y)$ be the rank of $y$ among all $Y_j$'s.
For a non-negative continuous bivariate function $h: [0,1]^2 \to [0,\infty)$ satisfying $h(x,x) =0$ for all $x \in [0,1]$, 
define the variation statistic
\begin{eqnarray} \label{eq_1.3+}
\tilde \zeta^{(h)}_n(X,Y) := \frac{1}{n}\sum_{i=1}^{n-1} h\Big( \frac{R(Y_{[i]})}{n}, \frac{R(Y_{[i+1]})}{n}\Big), 
\end{eqnarray} and the normalization statistic 
\begin{eqnarray}
\tilde \chi^{(h)}_n(Y) := \frac{1}{n^2}\sum_{i=1}^n \sum_{j=1}^n h\Big( \frac{R(Y_{i})}{n}, \frac{R(Y_{j})}{n}\Big). \label{eq_nor_q+}
\end{eqnarray}The proposed rank-based correlation coefficient is defined as
\begin{eqnarray*}
\tilde \xi^{(h)}_n(X,Y) &:=& 1 - \tilde \zeta^{(h)}_n(X,Y)/\tilde \chi^{(h)}_n(Y) \cr
&=& 1 - \frac{n \cdot\sum_{i=1}^{n-1} h\big(n^{-1}\cdot R(Y_{[i]}), n^{-1} \cdot R(Y_{[i+1]})\big)}{\sum_{i=1}^n \sum_{j=1}^n h\big(n^{-1} \cdot R(Y_{i}), n^{-1} \cdot R(Y_{j})\big)}.
\end{eqnarray*}
In case that $\tilde \chi^{(h)}_n(Y) =0 $, we set $\tilde \xi^{(h)}_n(X,Y) = 1$. 
\end{definition}

It is noteworthy that all the theorems in previous sections are derived for $\xi^{(h,F)}_n(X,Y)$ with deterministic $h$ and $F$, while the properties of the rank-based coefficient $\tilde \xi^{(h)}_n(X,Y) = \xi^{(h,F_{Y,n})}_n(X,Y)$ with non-deterministic $F_{Y,n}$ remains unexplored. 
In our context, it is important to note that the empirical $\cdf$ $F_{Y,n}$ differs fundamentally from a typical $\cdf$ $F$ in Definition \ref{Def_1}.
First, from an asymptotic standpoint, $F_{Y,n}$ is a random function depending on the sample $Y_i$'s and sample size $n$, whereas $F$ is deterministic and independent from $Y$. This disparity significantly impacts the derivation of asymptotic theories. Second, $F_{Y,n}$ in its nature is a piecewise-constant function, which obviously violates the continuity assumption for $F$ in condition \ref{CF1}. For these reasons, the asymptotic properties of $\tilde \xi^{(h)}_n(X,Y)$ can not be directly inferred from the previous theorems, but need to be investigated separately and specifically.

Theorem \ref{Thm_5} shows that $\tilde \xi^{(h)}_n(X,Y)$ also converges almost surely to a deterministic limit, provided that function $h$ is Lipschitz-continuous. 

\begin{enumerate}[label=\rm{(C\arabic*)}]
\addtocounter{enumi}{5}  
\item (Lipschitz continuity condition on $h$). \label{Ch4}
There exists a constant $K_h\in (0,\infty)$, such that 
$|h(y_1,z_1) - h(y_2,z_2)| \leq K_h \cdot (|y_1-y_2| + |z_1-z_2|)$,
for any $(y_1, z_1), (y_2, z_2) \in [0,1]^2$.
\end{enumerate}
\begin{theorem}[Convergence of rank-based correlation coefficient] \label{Thm_5}
Assume that $h$ satisfies conditions \ref{Ch1} and \ref{Ch4}.
As $n \to \infty$, we have the following three convergence results: $\mathrm{(i)}$ the variation statistic $\tilde \zeta^{(h,F)}_n(X,Y)$ converges a.s. to the limit
\begin{eqnarray*}
\tilde \zeta^{(h)} &:=& \int\int\int h\big(F_Y(y),F_Y(z)\big)\ \d \mu_{Y|x}(y) \ \d \mu_{Y|x}(z) \ \d \mu_{X}(x);  
\end{eqnarray*}
$\mathrm{(ii)}$ the normalization statistic $\tilde \chi^{(h)}_n(Y) $ converges a.s. to the strictly-positive limit
\begin{eqnarray}
\tilde \chi^{(h)} := \int\int h(F_Y(y),F_Y(z))\ \d \mu_{Y}(y) \ \d \mu_{Y}(z); \label{eq: chi++}
\end{eqnarray}
and $\mathrm{(iii)}$ the correlation coefficient $\tilde \xi^{(h)}_n(X,Y)$ converges a.s. to the limit 
\begin{eqnarray}
\tilde \xi^{(h)} &:=& 1- \tilde \zeta^{(h)} /\tilde \chi^{(h)}.     \label{eq: xi++}
\end{eqnarray}
\end{theorem}

Moreover, when $X$ and $Y$ are independent, the central limit theorem also holds for $\tilde \xi^{(h)}_n(X,Y)$.
\begin{theorem}[CLT of $\tilde \xi^{(h)}_n(X,Y)$ under independence]\label{Thm_6}
Assume that $h$ satisfies conditions \ref{Ch1} and \ref{Ch4}. 
Assume that $X$ and $Y$ are independent. Then as $n \to \infty$, we have
\begin{eqnarray*}
\sqrt{n} \cdot \tilde \xi^{(h)}_n(X,Y) \conD N(0, \tilde \sigma^2), 
\end{eqnarray*}
where $\tilde \sigma^2$ 
is obtained by plugging $F = F_{Y}$ into the expression of $\sigma^2$ in Theorem \ref{Thm_3}.
\end{theorem}

\begin{remark} \label{remark_4.1}
It is a well-known fact that $F_Y(Y)$ follows a Unif$[0,1]$ distribution if $Y$ is continuous. 
In such case, $\tilde \sigma^2$ reduces to $\big\{\E h(U_1,U_2)^2 - 2\cdot\E \big( h(U_1, U_2)\cdot h(U_1,U_3) \big) +\big(\E h(U_1,U_2)\big)^2\}/\big(\E h(U_1,U_2)\big)^2$ with $U_1, U_2, U_3 \sim_\iid \text{Unif}[0,1]$. Owing to this expression, $\tilde \sigma^2$ can be directly computed via numerical integrals, without the need of being estimated from the observed data.  For example, when taking $h(y,z) = |y-z|^\gamma$ with $\gamma >0$, we have
\begin{eqnarray*}
\tilde \sigma^2 = 1+(\gamma+2)^2 \cdot\Big\{
\frac{\gamma+1}{4(2\gamma+1)}
-
\frac{1}{2\gamma+3}
-
\frac{\Gamma(\gamma+2)^2}{\Gamma(2\gamma+4)}
\Big\},
\end{eqnarray*}
where $\Gamma(\cdot)$ represents the gamma function. Letting $\gamma =1$ yields $\tilde \sigma^2  = 2/5$, which agrees with the variance $\tau^2$ in \cite{Chatterjee_2021} [Theorem 2.2].
\end{remark}

\subsection{Simplified rank-based coefficient under continuous $Y$}

Note that $F_Y(Y) \sim \text{Unif}[0,1]$ for any continuous random variable $Y$. Therefore, if $Y$ is continuous, the denominator term $\tilde \chi^{(h)}$ in \eqref{eq: chi++} could be simplified as
\begin{eqnarray}
\tilde \chi^{(h)} = \int_{0}^1\int_0^1 h(u,v)\ \d u \ \d v =: C_h, \label{eq: C_h}
\end{eqnarray}
which is a positive constant that does not rely on the distribution law of $Y$. In this scenario, the normalization statistic $\tilde \chi^{(h)}_n(Y)$ in \eqref{eq_nor_q+}, acting as consistent estimator for $\tilde \chi^{(h)}$, could be simply replaced by the exact value $C_h = \tilde \chi^{(h)}$. This gives rises to a simplified version of the rank-based correlation coefficient, as presented in Definition \ref{Def_3}.

\begin{definition}[Simplified rank-based correlation coefficient under continuous $Y$] \label{Def_3}
Assume that $Y$ is continuous. For a non-negative non-constant continuous bivariate function $h: [0,1]^2 \to [0,\infty)$ satisfying $h(x,x) =0$ for all $x \in [0,1]$, define the simplified rank-based correlation coefficient 
\begin{eqnarray*}
\breve \xi^{(h)}_n(X,Y) &:=& 1 - \tilde \zeta^{(h)}_n(X,Y)/C_h \cr
&=& 1 - \frac{n^{-1}\cdot\sum_{i=1}^{n-1} h\big(n^{-1} \cdot R(Y_{[i]}), n^{-1} \cdot R(Y_{[i+1]})\big)}{ \int_{0}^1\int_0^1 h(u,v)\ \d u \ \d v},
\end{eqnarray*}
where $\tilde \zeta^{(h)}_n(X,Y)$ and $C_h$ are defined in \eqref{eq_1.3+} and \eqref{eq: C_h} respectively. 
\end{definition}

A major appealing property for  $\breve \xi^{(h)}_n(X,Y)$ is its reduced computational cost compared to that of the original $\tilde \xi^{(h)}_n(X,Y)$. This is because the normalization constant $C_h$ in $\breve \xi^{(h)}_n(X,Y)$ does not rely on $Y$ and could be directly computed from the integral  \eqref{eq: C_h}. 
In fact, for many common $h$ functions, $C_h$ has simple close-form expressions. Take our previous examples of $h$ in Section \ref{Sec_3.2}, if $h(y,z) = |y-z|^\gamma$ for $\gamma >0$, then $C_h = 2\cdot (\gamma+1)^{-1} \cdot (\gamma+2)^{-1}$; and if $h(y,z) = 1-\exp(-\beta \cdot |y-z|)$ for $\beta >0$, then $C_h = 1-2 \beta^{-1} + 2 \beta^{-2} -2  \beta^{-2} \cdot\exp(-\beta)$. Without the need of evaluating the normalization statistic, the computational complexity of $\breve \xi^{(h)}_n(X,Y)$ is reduced from $O(n^2)$ to $O(n \log n)$. 
Corollary \ref{Corr_2} ensures that $\breve \xi^{(h)}_n(X,Y)$ preserves all the asymptotic properties of $\tilde \xi^{(h)}_n(X,Y)$ in Theorems \ref{Thm_5} and \ref{Thm_6}.

\begin{corollary}[Consistency and CLT for simplified correlation coefficient] \label{Corr_2}
Assume $Y$ is continuous. 	Assume $h$ satisfies conditions  
\ref{Ch1} and \ref{Ch4}.
Then as $n \to \infty$, $\breve \xi^{(h)}_n(X,Y)$ converges a.s. to the limit $\tilde \xi^{(h)}$ in \eqref{eq: xi++}. Moreover, if $X$ and $Y$ are independent, then the central limit theorem holds: 
$\sqrt{n} \cdot \breve \xi^{(h)}_n(X,Y) \conD N(0, \tilde \sigma^2)$ as $n \to \infty$, where $ \tilde \sigma^2$ is defined in Theorem \ref{Thm_6}. 

\end{corollary}

To conclude, $\breve \xi^{(h)}_n(X,Y)$ achieves the same nice asymptotic properties as  $\tilde \xi^{(h)}_n(X,Y)$, but
with much reduced computational expense. Therefore, $\breve \xi^{(h)}_n(X,Y)$ is a more efficient alternative to $\tilde \xi^{(h)}_n(X,Y)$, if knowing that $Y_i$'s take values in continuous support.

\subsection{Properties of correlation measure $\tilde \xi^{(h)}$}
This subsection investigates the properties of correlation measure $\tilde \xi^{(h)}$, which is the limit of the rank-based correlation coefficient $\tilde \xi^{(h)}_n(X,Y)$ or its simplified version $\breve \xi^{(h)}_n(X,Y)$.
A comparison between the correlation measures  $\tilde \xi^{(h)}$ in \eqref{eq: xi++} and $\xi^{(h,F)}$ in \eqref{eq: xi} gives that $\tilde \xi^{(h)}= \xi^{(h,F_Y)}$, that is, $\tilde \xi^{(h)}$ coincides with  $\xi^{(h,F)}$ when substituting $F = F_Y$. One may speculate that $\tilde \xi^{(h)}$ also preserves similar properties of $\xi^{(h,F)}$ as presented in Theorems \ref{Thm_2}, \ref{Thm_4}, and Proposition \ref{Prop_1}. Again, this needs to be reexamined, as we should not simply regard $\tilde \xi^{(h)}= \xi^{(h,F_Y)}$ as a special case of $\xi^{(h,F)}$. For all previous theorems related to $\xi^{(h,F)}$, $F$ is assumed to be a fixed function that does not depend on the sample distribution and satisfies the regularity condition \ref{CF1}. In contrast, $F_Y$ not only depends on the sample distribution, but also possibly violates the required regularity condition \ref{CF1}. We stress that our investigation into properties \ref{CP_1}--\ref{CP_3} encompasses all possible probability distributions of $(X,Y)$, only except for constant $X$ or $Y$.

To proceed, we first present conditions \ref{Cphi1+} and \ref{Ch3+}, which are modified version of conditions \ref{Cphi1} and \ref{Ch3} in Section \ref{Sec_3.2}.

\begin{enumerate}[label=\rm{(C\arabic*)}$'$]
\addtocounter{enumi}{3}  
\item \label{Cphi1+} 
(Modified condition on $\varphi$). $\varphi: [0,1]^2 \to \mathbbR$ is a continuous positive-definite symmetric function, such that for any two distinct probability measures $\mu_1$ and $\mu_2$ satisfying $\mu_1((0,1]) = \mu_2((0,1])=1$, the set
$\calT_{\mu_1, \mu_2} := \{t \in [0,1]: \E_{Z\sim \mu_1} \varphi(t, Z) \neq \E_{Z\sim \mu_2} \varphi(t, Z) \}$ 	is non-empty. 
\end{enumerate}

\begin{enumerate}[label=\rm{(C\arabic*)}$'$]
\addtocounter{enumi}{4}  
\item (Modified equivalence condition on $h$ for achieving \ref{CP_1}--\ref{CP_3}). \label{Ch3+} 
There exists some bivariate function $\varphi: [0,1]^2 \to \mathbbR$ satisfying condition \ref{Cphi1+}, such that $h(y,z) = 1/2\cdot \{\varphi(y,y)+\varphi(z,z)\} -  \varphi(y,z)$ holds for all $y,z \in [0,1]$. 
\end{enumerate}
Note that the difference between conditions  \ref{Cphi1+} and  \ref{Cphi1} is minor: the open interval $(0,1)$ in condition \ref{Cphi1} is modified to be the wider half-open interval $(0,1]$ in condition \ref{Cphi1} (i.e., ``$\mu_1((0,1)) = \mu_2((0,1))=1$'' in \ref{Cphi1} is revised to be ``$\mu_1((0,1]) = \mu_2((0,1])=1$'' in \ref{Cphi1+}). Therefore, the revised conditions \ref{Cphi1+} and \ref{Ch3+} are slightly stronger than the previous conditions  \ref{Cphi1} and \ref{Ch3}. Theorem \ref{Thm_7} verifies that $\tilde \xi^{(h)}$  possesses similar properties to $\xi^{(h,F)}$. 
\begin{theorem} [Properties of correlation measure $\tilde \xi^{(h)}$] \label{Thm_7}
Assume that $h$ satisfies the regularity condition \ref{Ch1}. Then we have the following results. 
\begin{enumerate}[label=\rm{(\roman*)}]
\item $\tilde \xi^{(h)}$ possesses properties \ref{CP_2} and \ref{CP_3+}.
\item $\tilde \xi^{(h)}$ possesses properties \ref{CP_1}, \ref{CP_2} and \ref{CP_3+}, if and only if $h$ satisfies condition \ref{Ch2}.
\item $\tilde \xi^{(h)}$ possesses properties \ref{CP_1}, \ref{CP_2} and \ref{CP_3}, if $h$ satisfies condition \ref{Ch3+}.
\end{enumerate}
\end{theorem}

In comparison with the previous Corollary \ref{Coro_1} for $\xi^{(h,F)}$, most results in Theorem \ref{Thm_7} for $\tilde \xi^{(h)}$ remain unchanged. The only exception is that we solely confirmed condition \ref{Ch3+} as the sufficiency condition for attaining all properties \ref{CP_1}--\ref{CP_3}, leaving uncertain whether it is also the  necessary condition. This poses an open question for future resolution.

\subsection{Other related works}
There are several recent works that extended the Chatterjee's correlation coefficient $\xi_n$ in different directions. \cite{azadkia2021simple} proposed a correlation coefficient based on $\xi_n$ for measuring the conditional dependence of $Y$ on $Z$ given $X$, and accordingly designed a new variable selection algorithm. \cite{Lin_Han_2022_CLT} proved the CLT for Chatterjee's correlation coefficient under arbitrary distributions of $(X,Y)$.  \cite{auddy2021exact}, \cite{Bickel_Peter_2022} and \cite{Shi_Drton_Han_2021_Biometrika} conducted power analyses for $\xi_n$ under a family of alternatives.  \cite{Lin_Han_2022_Biometrika} extended $\xi_n$ to enhance power for independence tests, by incorporating multiple nearby ranks. Generalizations of $\xi_n$ to multi-dimensional spaces were proposed in \cite{huang2022kernel}, \cite{Gamboa_etal_2022_Bernoulli}, and \cite{Fuchs_2023_JOMA}. Scientific applications of $\xi_n$ across various fields could be found in \cite{Sadeghi_Behnam_2022}, \cite{Zarei_etal_2023}, and \cite{Nurdiati_etal_2022}. Refer to \cite{Chtterjee_2022_survey} for a comprehensive review of recent follow-up works. 
While these works focus on various topics such as independence test, power analysis, variable selection, and high-dimensional extensions, our research takes a distinct approach. We concentrate on tackling the fundamental task of quantifying functional dependence between two random scalars --- an objective that aligns with the original intention of the Chatterjee’s work. In pursuit of this goal, we have developed enhanced alternatives to the Chatterjee’s correlation coefficient.
As we will later show in numerical examples in Section \ref{Sec_5}, with appropriate choices of $h$ and $F$ functions, our newly proposed correlation coefficients 
are more sensitive in identifying perfect or near-perfect functional dependence than the Chatterjee's correlation coefficient. 

\section{Simulation studies}  \label{Sec_5}
In this section, we evaluate the performance of our proposed correlation coefficients via simulations on synthetic data. 
\subsection{Set-up}
For generating the synthetic data $\{(X_i,Y_i)\}_{i=1,\dots,n}$, we adopt the following three models as used in section 4 of \cite{Chatterjee_2021}:
\begin{description}
\item[Model 1 (Linear): ]  $Y = X + \sigma E.$
\item[Model 2 (Quadratic): ]  $Y = X^2 + \sigma E.$
\item [Model 3 (Sinusoidal): ] $Y =\sin(2\pi\cdot X) + \sigma E.$
\end{description}
In each of the three models mentioned above, we set $X \sim \text{Unif}[-1,1]$. 
Additionally, 
$E \sim N(0,1)$ is the Gaussian noise independent of $X$, and $\sigma \in [0,\infty)$ represents the noise magnitude.
With a slight abuse of notation, we also allow taking $\sigma = \infty$, 
which corresponds to the scenario where $Y = E$ is pure noise. 

We access the performance of the correlation coefficient $\xi_{n}^{(h,F)}$ in Definition \ref{Def_1} with $F = \Phi$ being the $\cdf$ of $N(0,1)$ distribution, as well as the simplified rank-based correlation coefficient $\breve \xi_{n}^{(h)}$ in Definition \ref{Def_3}.
For both of  $\xi_{n}^{(h,F)}$ and $\breve \xi_{n}^{(h)}$,
the bivariate function $h$ is specified from the following five cases: (1) $h_1(y,z) = |y-z|$; (2) $h_2(y,z)= (y-z)^2$; (3) $h_3(y,z) = |y-z|^3$; (4) $h_4(y,z) = (e^y - e^z)^2$; and (5) $h_5(y,z) = 1-e^{|y-z|}$. According to Corollary \ref{Coro_1} and Theorem \ref{Thm_7}, the correlation measures $\xi^{(h,F)}$ and $\tilde \xi^{(h)}$ can possess different properties of \ref{CP_1}--\ref{CP_3} contingent upon the specific $h$ functions. Specifically, $h_1$ and $h_5$ lead to properties \ref{CP_1}, \ref{CP_2}, and \ref{CP_3}; $h_2$ and $h_4$ result in properties \ref{CP_1}, \ref{CP_2}, and \ref{CP_3+}; and $h_3$ yields properties \ref{CP_2} and \ref{CP_3+}. These $h$ functions all give rise to the perfect dependence property \ref{CP_2}, indicating the validity of the corresponding correlation coefficients for detecting arbitrarily-shaped functional dependence. 

\begin{figure}[!htbp]
\centering
\raisebox{1 em}{
\includegraphics[width=\linewidth]
{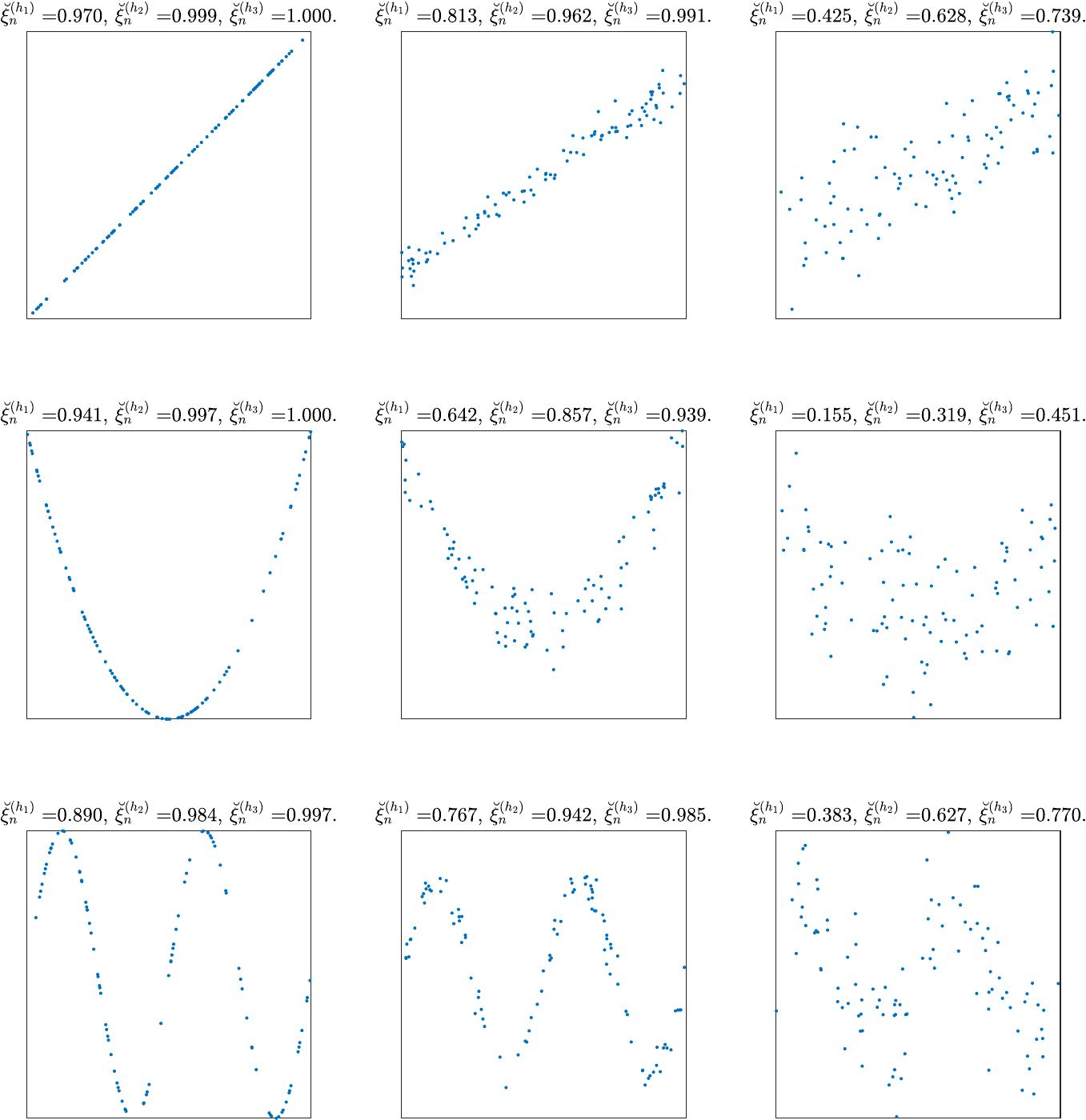}}
\caption{
Illustrative example for the simplified rank-based correlation coefficients $\breve \xi_{n}^{(h)}$, with $h_1 = |y-z|$, $h_2 = (y-z)^2$ and $h_3 = |y-z|^3$. In the 3-by-3 panel grids, rows from top to bottom correspond to Models 1, 2, and 3; and columns from left to right correspond to noise level $\sigma = 0, 0.1$ and $0.5$. The sample size $n = 100$. Note that $\breve \xi_{n}^{(h_1)}$ coincides with the Chatterjee's correlation coefficient. 
}
\label{Figure_1}
\end{figure}

\subsection{Illustrative example}
Figure \ref{Figure_1} provides an illustrative example on the performance of simplified rank-based correlation coefficient $\breve \xi_{n}^{(h)}$, with $h = h_1$, $h_2$ or $h_3$. Note that $\breve \xi_n^{(h)}$ with $h_1 = |y-z|$ coincides with the Chatterjee's correlation coefficient in \eqref{eq:CCC}, which serves as a benchmark for method comparisons. 
Across all models, the three correlation coefficients exhibit higher values when the noise level is lower.
In each panel of Figure \ref{Figure_1}, the three correlation coefficients consistently follow this order: $\breve \xi_{n}^{(h_1)} < \breve \xi_{n}^{(h_2)} < \breve \xi_{n}^{(h_3)}$. This suggests that $\breve \xi_{n}^{(h_2)}$ and $\breve \xi_{n}^{(h_3)}$ are more sensitive than $\breve \xi_{n}^{(h_1)}$ in identifying functional dependence, especially under the case of 
winding-shaped Models 2 and 3. In particular, for the noise-free case $\sigma = 0$,  all three correlation coefficients are expected to converge to $1$ according to the perfect dependence property \ref{CP_2}. 
It is observed that $\breve \xi_{n}^{(h_2)}$ and $\breve \xi_{n}^{(h_3)}$ are much closer to $1$ (with the largest difference $< 0.02$) than $\breve \xi_{n}^{(h_1)}$ (with the largest difference $> 0.1$). 
This indicates that $\breve \xi_{n}^{(h_2)}$ and $\breve \xi_{n}^{(h_3)}$ have higher convergence speed than $\breve \xi_{n}^{(h_1)}$, or in other words,
$\breve \xi_{n}^{(h_2)}$ and $\breve \xi_{n}^{(h_3)}$ can attain enhanced detection results with a reduced sample size requirement.
This simulation example reveals that among our proposed family of correlation coefficients, some $\breve \xi_{n}^{(h)}$ with appropriately-chosen $h$ functions may have better practical abilities for detecting functional relationship than the benchmark Chatterjee's correlation coefficient.

\subsection{Comparison with other correlation coefficients}

To provide a more comprehensive evaluation, we compare our correlation coefficient with some other methods, which includes the Peason's correlation coefficient, the Spearman's $\rho$, the distance correlation (abbreviated as ``dCorr'') in \cite{Szekely_etal_2007_AOS}, the power-enhanced Chatterjee's correlation coefficient in \cite{Lin_Han_2022_Biometrika} (abbreviated as ``LH'', with the number of nearest neighbor $M = 10$), and the Maximal Information Coefficient (MIC) in \cite{Reshef_etal_2011_Science}. 

Under the linear Model 1, all methods exhibit similar performances; therefore, the results are omitted.
Tables \ref{Table_1} and \ref{Table_2} report the values of correlation coefficients of each method under Models 2 and 3 respectively. 
As the sample size $n$ grows from $100$ to $2000$, the mean values of each correlation coefficient slightly changes, whereas the standard deviations become closer to zero. This implies a trend of asymptotic convergence for $\xi_{n}^{(h,F)}$ and $\breve \xi_n^{(h)}$, which agrees with our Theorems \ref{Thm_1} and \ref{Thm_5}.
Under both the two models, our proposed correlation coefficients $\xi_{n}^{(h,F)}$ and $\breve \xi_n^{(h)}$ distinguish themselves from other methods by clearly reflecting the amount of functional dependence between $X$ and $Y$. 
It is seen that $\xi_{n}^{(h,F)}$ and $\breve \xi_n^{(h)}$ are approximately equal to $1$ under perfect dependence (i.e., $\sigma = 0$), then decrease as the noise level $\sigma$ grows, and finally approaches to zero when $Y = E$ (i.e., $\sigma = \infty$) consists of pure noise. The MIC shows a similar trend, but it does not approach zero closely when the noise level goes to infinity, as it has the largest value among all methods under the case of $\sigma = \infty$.
Though the MIC is theoretically guaranteed to converge to zero (i.e., satisfying property \ref{CP_3+}) for independent $X$ and $Y$, our simulation results indicate that its convergence speed is significantly slower than other methods. 
The LH correlation coefficient is comparable to the Chatterjee's correlation coefficient $\breve \xi_n^{(h_1)}$ when the sample size is large ($n \in \{500,2000\}$), but exhibits inferior performance when the sample size is small ($n=100$). The distance correlation ``dCorr'' shows a trend of decreasing to zero as $\sigma$ grows, while it is far away from reaching $1$ for the noise-free case ($\sigma = 0$). 
The Pearson correlation coefficient and Spearman's $\rho$ fail to identify significant functional relationships in both models, as they are either negative or close to zero even under strong functional dependence ($\sigma \in \{0,0.1\}$).
For the comparison between the two proposed coefficient $\xi_{n}^{(h,F)}$ and $\breve \xi_n^{(h)}$, 
they exhibit similar values when using identical $h$ functions.
Among the five selected $h$ functions, $h_2$, $h_3$ and $h_4$ yield higher values of correlation coefficients than $h_1$ and $h_5$ do, especially when the functional dependence is strong ($\sigma \in \{0,0.1,0.5\}$). This also agrees with the previous results reflected in Figure \ref{Figure_1}.

\begin{table}[!htbp]
\caption{
Simulation results under Model 2: $Y = X^2 + \sigma E$.
Experiments are replicated for $100$ times, with mean values of correlation coefficients shown in the table, and 
standard deviations (multiplied by $100$) presented in parentheses.  For  $\xi_n^{(h,F)}$, we use $F = \Phi$, the $\cdf$ of standard normal distribution. 
}
\label{Table_1}
\begin{center}
\resizebox{\linewidth}{!}{
\begin{tabular}{ccllllllllllll}\toprule\multicolumn{2}{c}{\multirow{2}[2]{*}{Method}} & \multicolumn{3}{c}{$\sigma = 0, \ (Y = f(X))$} & \multicolumn{3}{c}{$\sigma = 0.1$} & \multicolumn{3}{c}{$\sigma = 0.5$} & \multicolumn{3}{c}{$\sigma = \infty, \  (X \perp Y)$ } \\
	\cmidrule(lr){3-5}  \cmidrule(lr){6-8}
	\cmidrule(lr){9-11}  \cmidrule(lr){12-14}
	\multicolumn{2}{c}{} & \multicolumn{1}{c}{n=100} & \multicolumn{1}{c}{n=500} & \multicolumn{1}{c}{n=2000} & \multicolumn{1}{c}{n=100} & \multicolumn{1}{c}{n=500} & \multicolumn{1}{c}{n=2000} & \multicolumn{1}{c}{n=100} & \multicolumn{1}{c}{n=500} & \multicolumn{1}{c}{n=2000} & \multicolumn{1}{c}{n=100} & \multicolumn{1}{c}{n=500} & \multicolumn{1}{c}{n=2000} \\
	\midrule\multicolumn{1}{r}{\multirow{10}[2]{*}{$\xi_n^{(h,F)}$}} & \multicolumn{1}{l}{$h_1= |y-z|$} & 0.943 & 0.989 & 0.997 & 0.674 & 0.680 & 0.680 & 0.126 & 0.145 & 0.143 & 0.004 & -0.003 & -0.001 \\      &       & (0.25) & (0.03) & (0.00) & (3.20) & (1.66) & (0.92) & (6.26) & (2.94) & (1.92) & (6.16) & (2.62) & (1.72) \\      & \multicolumn{1}{l}{$h_2 = (y-z)^2$} & 0.995 & 1.000 & 1.000 & 0.890 & 0.893 & 0.893 & 0.221 & 0.251 & 0.248 & -0.000 & -0.003 & -0.002 \\      &       & (0.08) & (0.00) & (0.00) & (2.00) & (1.03) & (0.57) & (10.01) & (4.49) & (2.97) & (9.57) & (4.07) & (2.88) \\      & \multicolumn{1}{l}{$h_3= |y-z|^3$} & 0.999 & 1.000 & 1.000 & 0.961 & 0.963 & 0.962 & 0.296 & 0.332 & 0.330 & -0.008 & -0.002 & -0.003 \\      &       & (0.03) & (0.00) & (0.00) & (1.04) & (0.54) & (0.29) & (13.22) & (5.81) & (3.73) & (12.90) & (5.53) & (3.88) \\      & \multicolumn{1}{l}{$h_4= (e^y-e^z)^2$} & 0.994 & 1.000 & 1.000 & 0.897 & 0.902 & 0.901 & 0.235 & 0.264 & 0.262 & -0.003 & -0.003 & -0.002 \\      &       & (0.10) & (0.00) & (0.00) & (1.90) & (0.93) & (0.51) & (10.15) & (4.46) & (2.91) & (9.65) & (4.20) & (2.93) \\      & \multicolumn{1}{l}{$h_5 = 1-e^{|y-z|}$} & 0.939 & 0.988 & 0.997 & 0.653 & 0.659 & 0.659 & 0.111 & 0.128 & 0.127 & 0.005 & -0.003 & -0.001 \\      &       & (0.25) & (0.03) & (0.00) & (3.27) & (1.70) & (0.93) & (5.77) & (2.73) & (1.76) & (5.53) & (2.36) & (1.49) \\\midrule\multicolumn{1}{r}{\multirow{10}[2]{*}{$\breve \xi_n^{(h)}$}} & \multicolumn{1}{l}{$h_1 = |y-z|$} & 0.941 & 0.988 & 0.997 & 0.645 & 0.660 & 0.664 & 0.134 & 0.155 & 0.155 & 0.004 & -0.003 & 0.000 \\      & (Chatterjee) & (0.04) & (0.00) & (0.00) & (3.50) & (1.94) & (0.91) & (6.66) & (3.07) & (1.73) & (6.10) & (2.63) & (1.38) \\      & \multicolumn{1}{l}{$h_2 = (y-z)^2$} & 0.997 & 1.000 & 1.000 & 0.859 & 0.865 & 0.869 & 0.225 & 0.257 & 0.257 & -0.000 & -0.003 & 0.000 \\      &       & (0.03) & (0.00) & (0.00) & (2.90) & (1.59) & (0.71) & (10.27) & (4.47) & (2.54) & (9.52) & (4.08) & (2.16) \\      & \multicolumn{1}{l}{$h_3= |y-z|^3$} & 1.000 & 1.000 & 1.000 & 0.938 & 0.941 & 0.943 & 0.292 & 0.331 & 0.331 & -0.008 & -0.002 & -0.001 \\      &       & (0.01) & (0.00) & (0.00) & (2.04) & (1.08) & (0.47) & (12.85) & (5.48) & (3.05) & (12.84) & (5.54) & (2.84) \\      & \multicolumn{1}{l}{$h_4 = (e^y-e^z)^2$} & 0.996 & 1.000 & 1.000 & 0.816 & 0.822 & 0.827 & 0.200 & 0.239 & 0.241 & -0.008 & -0.004 & -0.000 \\      &       & (0.04) & (0.00) & (0.00) & (4.06) & (2.20) & (0.97) & (10.22) & (4.53) & (2.51) & (9.61) & (4.00) & (2.13) \\      & \multicolumn{1}{l}{$h_5= 1-e^{|y-z|}$} & 0.927 & 0.985 & 0.996 & 0.593 & 0.609 & 0.614 & 0.113 & 0.132 & 0.132 & 0.005 & -0.003 & 0.000 \\      &       & (0.05) & (0.00) & (0.00) & (3.65) & (2.02) & (0.96) & (5.93) & (2.79) & (1.56) & (5.48) & (2.36) & (1.23) \\\midrule\multicolumn{2}{c}{Pearson} & -0.005 & -0.007 & 0.003 & 0.021 & 0.007 & 0.005 & -0.004 & -0.006 & -0.004 & 0.003 & 0.007 & -0.000 \\\multicolumn{2}{c}{} & (12.98) & (5.04) & (2.42) & (12.10) & (5.83) & (2.72) & (11.08) & (4.51) & (2.39) & (8.65) & (4.75) & (2.33) \\\multicolumn{2}{c}{Spearman} & -0.003 & -0.007 & 0.000 & 0.010 & 0.002 & 0.005 & -0.004 & -0.007 & -0.003 & 0.003 & 0.009 & -0.001 \\\multicolumn{2}{c}{} & (14.46) & (5.64) & (2.67) & (12.91) & (5.96) & (3.27) & (12.02) & (4.81) & (2.38) & (8.63) & (4.65) & (2.41) \\\multicolumn{2}{c}{dCorr} & 0.514 & 0.496 & 0.492 & 0.485 & 0.468 & 0.462 & 0.278 & 0.245 & 0.239 & 0.160 & 0.074 & 0.038 \\\multicolumn{2}{c}{} & (2.82) & (1.04) & (0.46) & (2.65) & (1.05) & (0.53) & (4.37) & (1.75) & (0.91) & (3.15) & (1.68) & (0.90) \\\multicolumn{2}{c}{LH} & 0.611 & 0.918 & 0.979 & 0.470 & 0.638 & 0.660 & 0.026 & 0.134 & 0.150 & -0.161 & -0.032 & -0.008 \\\multicolumn{2}{c}{} & (0.00) & (0.00) & (0.00) & (2.45) & (1.59) & (0.75) & (5.18) & (2.05) & (1.28) & (2.62) & (0.95) & (0.46) \\\multicolumn{2}{c}{MIC} & 1.000 & 1.000 & 1.000 & 0.876 & 0.847 & 0.813 & 0.342 & 0.285 & 0.240 & 0.238 & 0.163 & 0.107 \\\multicolumn{2}{c}{} & (0.00) & (0.00) & (0.00) & (6.26) & (2.96) & (1.57) & (6.27) & (2.63) & (1.59) & (3.62) & (1.40) & (0.51) \\\bottomrule\end{tabular}%
}
\end{center}
\end{table}

\begin{table}[!htbp]
\caption{
Simulation results under Model 3: $Y = \sin(2\pi \cdot X) + \sigma E$.
Experiments are replicated for $100$ times, with mean values of correlation coefficients shown in the table, and 
standard deviations (multiplied by $100$) presented in parentheses.  For $\xi_n^{(h,F)}$, we use $F = \Phi$, the $\cdf$ of standard normal distribution. 
}
\label{Table_2}
\begin{center}
\resizebox{\linewidth}{!}{
\begin{tabular}{ccllllllllllll}\toprule\multicolumn{2}{c}{\multirow{2}[2]{*}{Method}} & \multicolumn{3}{c}{$\sigma = 0, \ (Y = f(X))$} & \multicolumn{3}{c}{$\sigma = 0.1$} & \multicolumn{3}{c}{$\sigma = 0.5$} & \multicolumn{3}{c}{$\sigma = \infty, \  (X \perp Y)$ } \\
	\cmidrule(lr){3-5}  \cmidrule(lr){6-8}
	\cmidrule(lr){9-11}  \cmidrule(lr){12-14}
	\multicolumn{2}{c}{} & \multicolumn{1}{c}{n=100} & \multicolumn{1}{c}{n=500} & \multicolumn{1}{c}{n=2000} & \multicolumn{1}{c}{n=100} & \multicolumn{1}{c}{n=500} & \multicolumn{1}{c}{n=2000} & \multicolumn{1}{c}{n=100} & \multicolumn{1}{c}{n=500} & \multicolumn{1}{c}{n=2000} & \multicolumn{1}{c}{n=100} & \multicolumn{1}{c}{n=500} & \multicolumn{1}{c}{n=2000} \\\midrule\multicolumn{1}{r}{\multirow{10}[2]{*}{$\xi_n^{(h,F)}$}} & \multicolumn{1}{l}{$h_1= |y-z|$} & 0.907 & 0.981 & 0.995 & 0.840 & 0.874 & 0.876 & 0.462 & 0.468 & 0.470 & 0.004 & -0.003 & -0.001 \\      &       & (0.37) & (0.03) & (0.01) & (1.13) & (0.56) & (0.36) & (5.37) & (2.50) & (1.41) & (6.16) & (2.62) & (1.72) \\      & \multicolumn{1}{l}{$h_2 = (y-z)^2$} & 0.985 & 0.999 & 1.000 & 0.970 & 0.983 & 0.984 & 0.686 & 0.693 & 0.696 & -0.000 & -0.003 & -0.002 \\      &       & (0.18) & (0.00) & (0.00) & (0.44) & (0.15) & (0.10) & (6.06) & (2.76) & (1.54) & (9.57) & (4.07) & (2.88) \\      & \multicolumn{1}{l}{$h_3= |y-z|^3$} & 0.997 & 1.000 & 1.000 & 0.993 & 0.997 & 0.998 & 0.803 & 0.809 & 0.812 & -0.008 & -0.002 & -0.003 \\      &       & (0.11) & (0.00) & (0.00) & (0.20) & (0.03) & (0.02) & (5.65) & (2.58) & (1.39) & (12.90) & (5.53) & (3.88) \\      & \multicolumn{1}{l}{$h_4= (e^y-e^z)^2$} & 0.985 & 0.999 & 1.000 & 0.969 & 0.982 & 0.983 & 0.677 & 0.685 & 0.687 & -0.003 & -0.003 & -0.002 \\      &       & (0.18) & (0.00) & (0.00) & (0.52) & (0.16) & (0.10) & (5.92) & (2.66) & (1.63) & (9.65) & (4.20) & (2.93) \\      & \multicolumn{1}{l}{$h_5 = 1-e^{|y-z|}$} & 0.890 & 0.977 & 0.994 & 0.811 & 0.850 & 0.852 & 0.411 & 0.417 & 0.419 & 0.005 & -0.003 & -0.001 \\      &       & (0.39) & (0.03) & (0.00) & (1.23) & (0.64) & (0.39) & (5.15) & (2.42) & (1.36) & (5.53) & (2.36) & (1.49) \\\midrule\multicolumn{1}{r}{\multirow{10}[2]{*}{$\breve \xi_n^{(h)}$}} & \multicolumn{1}{l}{$h_1 = |y-z|$} & 0.886 & 0.976 & 0.994 & 0.784 & 0.833 & 0.837 & 0.446 & 0.459 & 0.462 & 0.004 & -0.003 & 0.000 \\      & (Chatterjee) & (0.18) & (0.01) & (0.00) & (1.71) & (0.86) & (0.41) & (4.88) & (2.36) & (1.23) & (6.10) & (2.63) & (1.38) \\      & \multicolumn{1}{l}{$h_2 = (y-z)^2$} & 0.985 & 0.999 & 1.000 & 0.950 & 0.969 & 0.971 & 0.675 & 0.687 & 0.691 & -0.000 & -0.003 & 0.000 \\      &       & (0.13) & (0.00) & (0.00) & (0.80) & (0.32) & (0.16) & (5.81) & (2.69) & (1.35) & (9.52) & (4.08) & (2.16) \\      & \multicolumn{1}{l}{$h_3= |y-z|^3$} & 0.997 & 1.000 & 1.000 & 0.988 & 0.994 & 0.994 & 0.797 & 0.808 & 0.812 & -0.008 & -0.002 & -0.001 \\      &       & (0.07) & (0.00) & (0.00) & (0.32) & (0.11) & (0.05) & (5.58) & (2.59) & (1.26) & (12.84) & (5.54) & (2.84) \\      & \multicolumn{1}{l}{$h_4 = (e^y-e^z)^2$} & 0.983 & 0.999 & 1.000 & 0.944 & 0.965 & 0.966 & 0.663 & 0.678 & 0.681 & -0.008 & -0.004 & -0.000 \\      &       & (0.14) & (0.00) & (0.00) & (1.13) & (0.53) & (0.23) & (6.51) & (3.12) & (1.37) & (9.61) & (4.00) & (2.13) \\      & \multicolumn{1}{l}{$h_5= 1-e^{|y-z|}$} & 0.860 & 0.970 & 0.992 & 0.742 & 0.799 & 0.803 & 0.392 & 0.405 & 0.408 & 0.005 & -0.003 & 0.000 \\      &       & (0.22) & (0.01) & (0.00) & (1.94) & (0.99) & (0.48) & (4.70) & (2.29) & (1.21) & (5.48) & (2.36) & (1.23) \\\midrule\multicolumn{2}{c}{Pearson} & -0.380 & -0.385 & -0.391 & -0.381 & -0.380 & -0.387 & -0.311 & -0.316 & -0.318 & 0.003 & 0.007 & -0.000 \\\multicolumn{2}{c}{} & (7.67) & (3.44) & (1.69) & (7.38) & (3.35) & (1.68) & (8.08) & (3.40) & (1.93) & (8.65) & (4.75) & (2.33) \\\multicolumn{2}{c}{Spearman} & -0.363 & -0.369 & -0.376 & -0.365 & -0.366 & -0.375 & -0.317 & -0.326 & -0.326 & 0.003 & 0.009 & -0.001 \\\multicolumn{2}{c}{} & (8.71) & (3.98) & (1.96) & (8.16) & (4.00) & (1.85) & (9.08) & (3.74) & (2.09) & (8.63) & (4.65) & (2.41) \\\multicolumn{2}{c}{dCorr} & 0.440 & 0.431 & 0.432 & 0.442 & 0.426 & 0.428 & 0.371 & 0.355 & 0.354 & 0.160 & 0.074 & 0.038 \\\multicolumn{2}{c}{} & (4.90) & (2.29) & (1.16) & (4.56) & (2.27) & (1.19) & (5.24) & (2.26) & (1.42) & (3.15) & (1.68) & (0.90) \\\multicolumn{2}{c}{LH} & 0.269 & 0.826 & 0.955 & 0.262 & 0.750 & 0.823 & 0.128 & 0.411 & 0.451 & -0.161 & -0.032 & -0.008 \\\multicolumn{2}{c}{} & (2.19) & (0.18) & (0.02) & (2.41) & (0.63) & (0.33) & (3.70) & (1.74) & (0.98) & (2.62) & (0.95) & (0.46) \\\multicolumn{2}{c}{MIC} & 1.000 & 1.000 & 1.000 & 0.990 & 0.995 & 0.979 & 0.671 & 0.640 & 0.608 & 0.238 & 0.163 & 0.107 \\\multicolumn{2}{c}{} & (0.00) & (0.00) & (0.00) & (1.82) & (0.78) & (0.82) & (8.53) & (3.58) & (2.13) & (3.62) & (1.40) & (0.51) \\\bottomrule\end{tabular}%
}
\end{center}
\end{table}

\begin{figure}[!htbp]
\centering
\raisebox{1 em}{
\includegraphics[width=\linewidth]{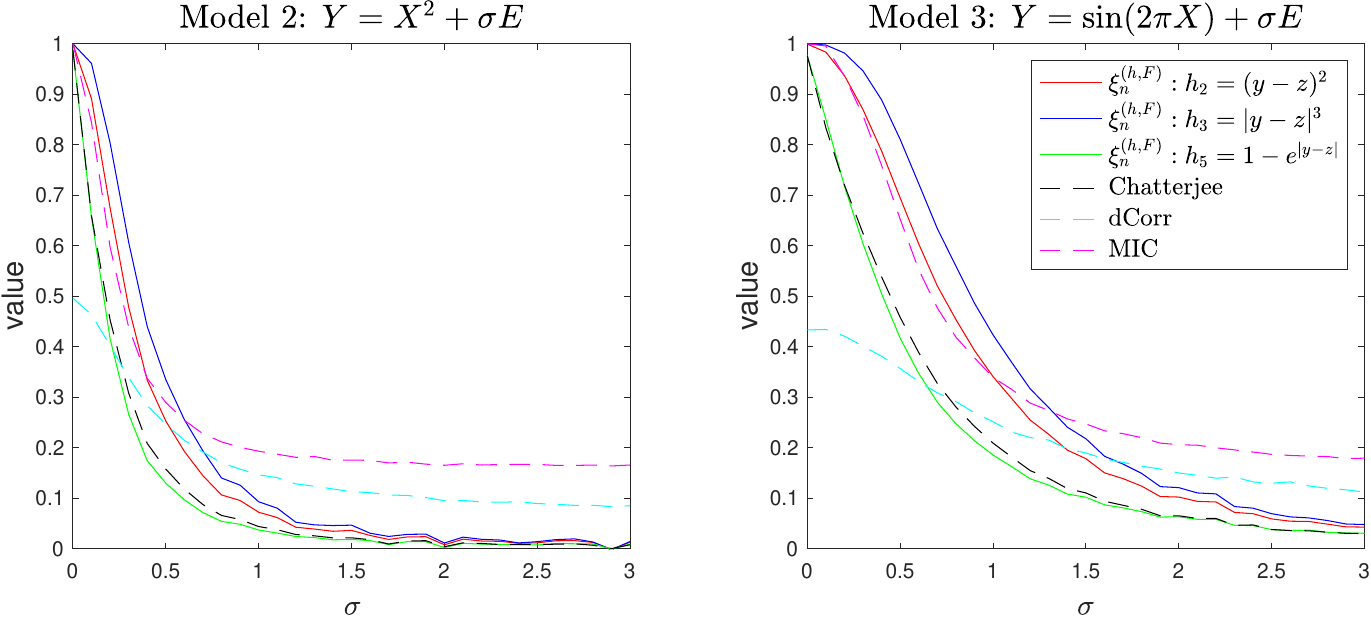}}
\caption{
Curves of selected correlation coefficients with varying $\sigma$.
The sample size is $n=500$. 
Experiments are replicated for $100$ times.  
}
\label{Figure_2}
\end{figure}

For a more intuitive presentation of results, Figure \ref{Figure_2} visualizes the trend of correlation coefficient with growing noise magnitude $\sigma$ for the six selected methods under Model 1 and Model 2. Among these methods, $\xi_{n}^{(h,F)}$ with $h = h_3$ proves to be the most desirable correlation coefficient under both models, as it exhibits the highest value in scenarios when the noise is small and dependence is strong, while rapidly approaching zero as $X$ and $Y$ become more independent.

\section{Real data experiment} \label{Sec_6}
The S\&P 500, also known as the Standard and Poor's 500, serves as a pivotal stock market index overseeing the performance of approximately 500 of the largest companies listed on United States stock exchanges. A primary objective for financial quant analysts involves examining the longitudinal trends and stability exhibited by these stock prices, which is crucial for refining effective trading strategies. Our proposed correlation coefficient is poised to meet this objective. Specifically, our correlation coefficient is intended for assessing the extent of functional dependence between each stock price $Y$ and the temporal variable $X$: higher dependence signifies a more stable trend in stock price, while lower dependence corresponds to augmented volatility and unpredictability within the stock. As an illustrative instance, our experiment scrutinizes the short-term volatility of the opening prices of the 504 stocks encompassed by the S\&P 500 index over a consecutive 50-day trading span from November 27, 2017, to February 7, 2018. The dataset is accessible at \url{https://www.kaggle.com/datasets/camnugent/sandp500}.

\begin{figure}[!htbp]
\centering
\raisebox{1 em}{
\includegraphics[width=\linewidth]{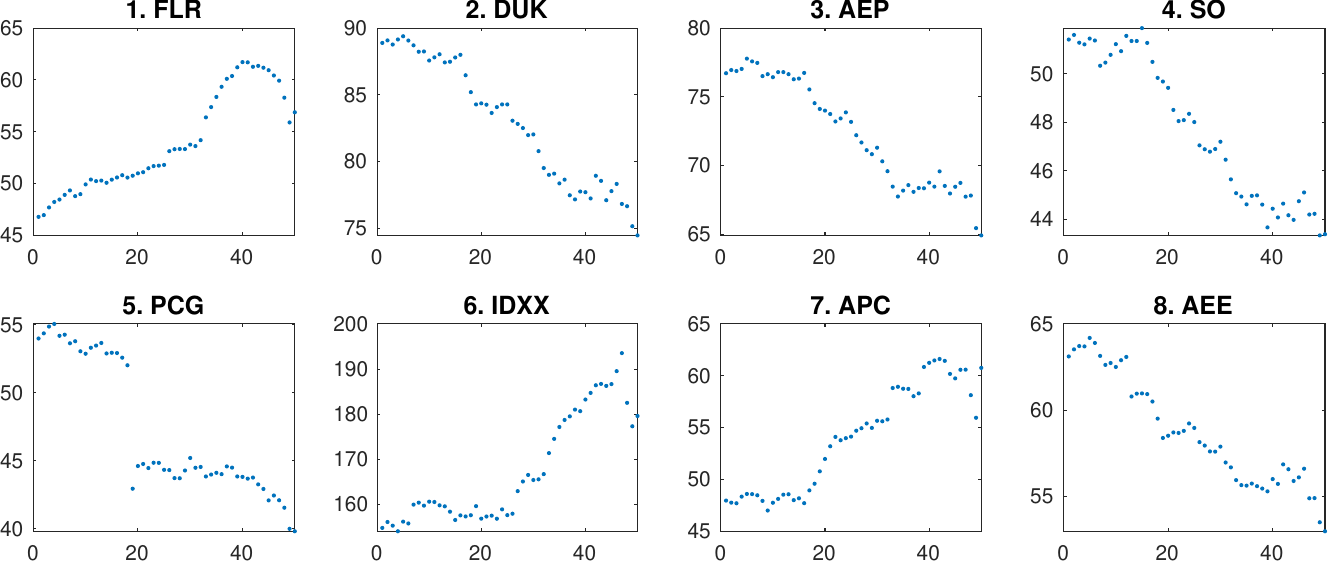}}
\caption{
S\&P 500 stock data: top 8 stocks picked by $\xi^{(h_1,F)}_n$. 
In each subplot, the horizontal axis represents the trading days (sequentially numbered from 1 to 50), and the vertical axis represents the stock price (measured in US dollars).
}
\label{Figure_3}
\end{figure}

\begin{figure}[!htbp]
\centering
\raisebox{1 em}{
\includegraphics[width=\linewidth]{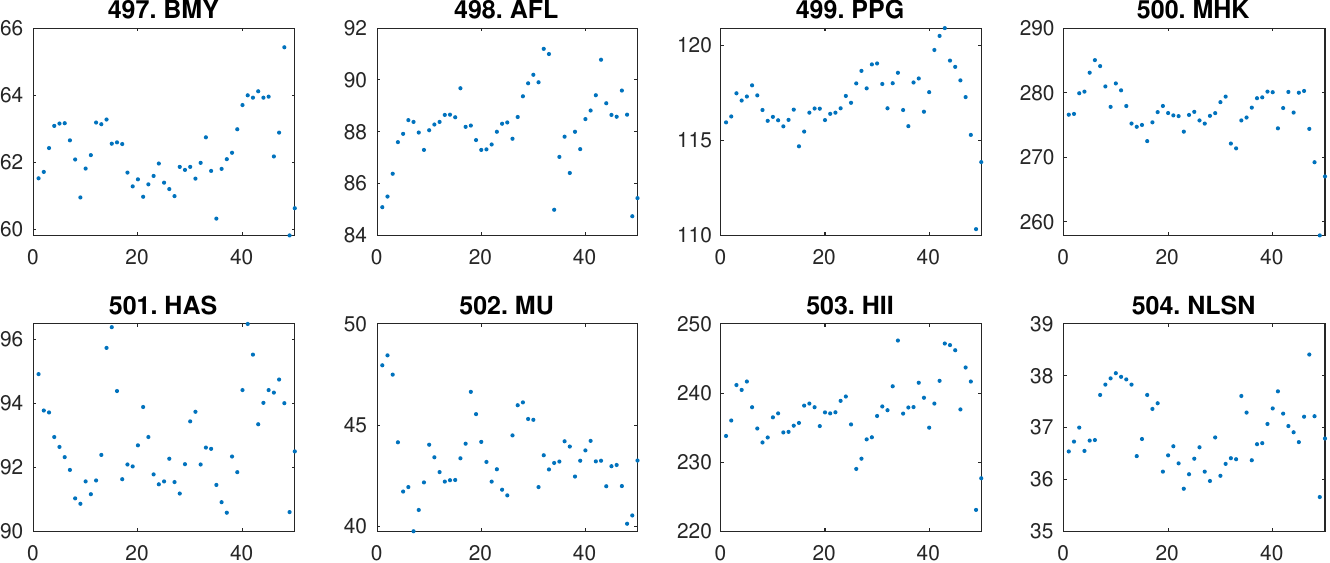}}
\caption{
S\&P 500 stock data: bottom 8 stocks picked by $\xi^{(h_1,F)}_n$.  In each subplot, the horizontal axis represents the trading days (sequentially numbered from 1 to 50), and the vertical axis represents the stock price (measured in US dollars).
}
\label{Figure_4}
\end{figure}

We first employ $\xi^{(h,F)}_n$ in \eqref{eq:our_CCC}, with $h$ function taking $h_1(y,z) = |y-z|$. 
For the $F$ function, we adopt $F(y) = \Phi(y-\mu)/\sigma$, where $\mu$ and $\sigma$ denote the mean and the standard deviation respectively of the observed sample $Y$,  and $\Phi(\cdot)$ is the $\cdf$ of standard normal distribution. To show the efficacy of our method, 
we sort these 504 stocks based on the values of their corresponding correlation coefficients  $\xi^{(h_1,F)}_n$. 
Figures \ref{Figure_3} and \ref{Figure_4} display scatter plots of the stock prices for the top 8 stocks with the highest $\xi^{(h_1,F)}_n$ values and the bottom 8 stocks with the lowest $\xi^{(h_1,F)}_n$ values, respectively.
In a stark contrast, the 8 stocks depicted in Figure \ref{Figure_3} exhibit smooth changes over time, whereas the 8 stocks in Figure \ref{Figure_4} demonstrate considerable fluctuations. 
Upon closer examination of the top 8 stocks in Figure \ref{Figure_3}, it is noticeable that certain stocks (e.g., DUK, AEP, PCG) follow a monotonic trend of changes, whereas others (e.g., FLR, IDXX, APC) exhibit smooth but relatively non-monotonic variations over time. 
This showcases the effectiveness of our correlation coefficient in identifying perfect functional dependence, regardless of whether this dependence is monotonic or not.

To draw a comparison, we also employ four other methods: our $\xi^{(h,F)}_n$ with $h$ function taking $h_5(y,z) = 1 - \exp(|y-z|)$, the Chatterjee's correlation coefficient, the MIC \citep{Reshef_etal_2011_Science}, and the distance correlation \citep{Szekely_etal_2007_AOS}. Table \ref{Table_12} displays the values of each correlation coefficient for the top 8 stocks (selected by $\xi^{(h_1,F)}_n$). 
The coefficient $\xi^{(h_5,F)}_n$ exhibits slight variation compared to $\xi^{(h_1,F)}_n$, yet maintains nearly identical rankings for the 8 selected stocks among the total 504 stocks. This suggests that transitioning from the $h_1$ function to $h_5$ in this context does not significantly alter the outcomes. The Chatterjee's correlation coefficient, also represented by $\breve \xi_n^{(h_1)}$ within our framework of simplified rank-based coefficient, displays more significant differences in rankings compared to $\xi^{(h_1,F)}_n$. 
This outcome is not surprising, given that a significant portion of the 504 stocks exhibit very high correlation coefficient values for both $\breve \xi_n^{(h_1)}$ and $\xi^{(h_1,F)}_n$. Consequently, even a slight alteration in coefficient values may result in substantial changes in their respective rankings among all stocks. For the MIC, it turns out that 177 out of the total 504 stocks share a MIC value of $1$ in common, and 7 out of the 8 selected stocks in Table \ref{Table_12} has MIC of $1$. Consequently, the MIC demonstrates limited utility as it fails to discern differences among many of these stocks. 
The distance correlation seems to be more sensitive at identifying monotonic relationships. However, for stocks like FLR and IDXX, which display non-monotonic functional dependence, the distance correlation tends to yield smaller values and lower ranks. Among the top 8 stocks in Figure \ref{Figure_3}, the fifth one PCG stands out notably. This is due to a significant price drop occurring around day 20, leading to the stock price's relationship with time resembling more of a piecewise continuous function. 
We anticipate that our proposed correlation coefficient is able to identify such ``piecewise smooth'' functional relationships. As indicated by property \ref{CP_2}, our correlation coefficient is expected to yield the highest value when $Y = f(X)$, where the function $f$ only needs to be measurable, but not necessarily to be continuous across the entire support of $X$. Consequently, such special type of ``piecewise smooth'' functional dependence is successfully identified by our $\xi^{(h_1,F)}_n$ and $\xi^{(h_5,F)}_n$, whereas it is overlooked by the other three approaches. In summary, this experiment highlights the effectiveness of our proposed correlation coefficient in real-world applications.

\begin{table}[!htbp]
\caption{
Values of selected correlation coefficients for the top 8 chosen stocks, along with their respective ranks among the total of 504 stocks. In $\xi_n^{(h,F)}$, we employ $h_1(y,z) = |y-z|$ and $h_5(y,z) = 1 - \exp(|y-z|)$.
}
\label{Table_12}
\begin{center}
\begin{tabular}{lcccccccccc}\toprule\multicolumn{1}{c}{\multirow{2}[2]{*}{Stock}} & \multicolumn{2}{c}{$\xi_n^{(h_1,F)}$ } & \multicolumn{2}{c}{$\xi_n^{(h_5,F)}$} & \multicolumn{2}{c}{Chatterjee } & \multicolumn{2}{c}{MIC} & \multicolumn{2}{c}{dCorr} \\   
\cmidrule(lr){2-3}  \cmidrule(lr){4-5}
\cmidrule(lr){6-7}  \cmidrule(lr){8-9}
\cmidrule(lr){10-11}
& value & rank  & value & rank  & value & rank  & value & rank  & value & rank \\\midrule FLR  & 0.920 & 1     & 0.900 & 1     & 0.885 & 1     & 1.000 & 1     & 0.944 & 27 \\ DUK  & 0.914 & 2     & 0.892 & 2     & 0.833 & 5     & 1.000 & 1     & 0.980 & 1 \\ AEP  & 0.904 & 3     & 0.878 & 4     & 0.784 & 57    & 1.000 & 1     & 0.972 & 2 \\ SO   & 0.902 & 4     & 0.876 & 5     & 0.772 & 92    & 1.000 & 1     & 0.971 & 6 \\ PCG  & 0.902 & 5     & 0.885 & 3     & 0.738 & 178   & 0.925 & 244   & 0.891 & 99 \\ IDXX & 0.896 & 6     & 0.869 & 8     & 0.792 & 44    & 1.000 & 1     & 0.914 & 62 \\ APC  & 0.896 & 7     & 0.871 & 6     & 0.803 & 29    & 1.000 & 1     & 0.960 & 16 \\ AEE  & 0.894 & 8     & 0.869 & 7     & 0.826 & 11    & 1.000 & 1     & 0.962 & 13 \\\bottomrule\end{tabular}%
\end{center}
\end{table}

\section{Discussion} \label{Sec_8}
Inspired by the utility of Chatterjee's correlation coefficient $\xi_n$ in measuring functional dependence,
we proposed three new types of correlation coefficients: the original proposal $ \xi^{(h,F)}_n$ in Definition \ref{Def_1}, the rank-based variant $\tilde \xi_n^{(h)}$ in Definition \ref{Def_2+} and its simplified version $\breve \xi_n^{(h)}$ in Definition \ref{Def_3}. 
We provide a systematic examination on the asymptotic properties of these coefficients, and demonstrate their capability for measuring functional dependence through both theoretical analysis and case studies. 

Several issues warrant further investigation.
First, how to select the the type of correlation coefficient $ \xi^{(h,F)}_n$, $\tilde \xi_n^{(h)}$, or  $\breve \xi_n^{(h)}$ in a specific real-world problem? 
From our theoretical results, $ \xi^{(h,F)}_n$, $\tilde \xi_n^{(h)}$, and  $\breve \xi_n^{(h)}$ have similar nice asymptotic properties; while in specific real applications, they may lead to notable differences. An example is that the two correlation coefficient $\xi^{(h_1,F)}_n$ and $\breve \xi^{(h_1)}_n$ (i.e., the Chatterjee's correlation coefficient), though with the identical $h$ functions, leads to notably different stock ranks in Table \ref{Table_12}. Conceptually, $\breve \xi^{(h)}_n$ is based on the ranks of $Y$, suggesting its greater robustness against outliers or abrupt changes; whereas $\xi^{(h,F)}_n$ is non-rank-based, enabling it to leverage more information from the distribution of $Y$ sample.
These distinct characteristics allow practitioners to select the type of correlation coefficient that best suits their specific requirements. Second, how to select $h$ and $F$ functions for our correlation coefficients? Again, our theoretical results provide guidances for selecting $h$ and $F$, e.g., we derive the necessary and sufficient condition on $h$ to achieve all desirable properties \ref{CP_1}--\ref{CP_3}. Yet the practical performance may differ with the theoretical expectations. For instance, $\xi^{(h_3,F)}_n$ with $h_3(y,z) = |y-z|^3$ only possesses part of the three properties \ref{CP_1}--\ref{CP_3}, but it demonstrates best practical performance in our simulation example, as reflected in Figure \ref{Figure_2}. In this regard, more practical guidance is needed for choosing $h$ and $F$. Thirdly, exploring further theoretical results, such as the convergence rate of correlation coefficients and the central limit theorem under non-independent $X$ and $Y$, holds potential for future investigation. 

\bibliographystyle{apalike}
\bibliography{References}

\end{document}